# Keeping Code-Aware LLMs Fresh: Full Refresh, In-Context Deltas, and Incremental Fine-Tuning

Authors: Pradeep Kumar Sharma, Ishaan Puri, Mantinder Jit Singh, Swapnil Shivaprasad, Hritvik Shrivastava

Affiliation: Persistent System

# Abstract

**Modern codebases evolve continuously: files are renamed or deleted; public APIs drift; behavior shifts within otherwise familiar modules. A model trained yesterday to map a developer's natural-language question to the exact set of repository file paths that matter will degrade tomorrow, even if the questions themselves look unchanged. In this paper we study—at system scale and across several widely used repositories—how to keep such a model fresh without surrendering retention on earlier code. We frame freshness as a form of domain drift between a base snapshot and the current HEAD, and we compare three families of update strategies: (A) Full Refresh, retraining the entire model at the new snapshot; (B) In-Context Learning (ICL) that injects recent deltas (raw git diffs or concise English summaries) at inference; and (C) Incremental Fine-Tuning (Inc-FT) on delta-derived training sets, with carefully controlled NEW:OLD mixing to mitigate catastrophic forgetting. We contribute an alias-aware evaluation protocol that credits rename while never rewarding deleted paths, and a practical Forgetting Probe that quantifies residual emissions of obsolete paths. Across Flask, SQL Alchemy, Pandas, and Poetry, Inc-FT with old-aware mixes delivers the best overall balance on mixed sets, ICL with English delta summaries delivers the fastest new-code lift when training is not feasible, and Full Refresh remains the ceiling when maximum NEW accuracy matters. We also compare Git-diff Inc-FT to Full-file Inc-FT, showing that diffs excel in rename/delete–heavy windows while full-file context wins in behavior-change–heavy windows.**

# Introduction

Modern engineering teams face a deceptively simple, relentlessly recurrent question: *given this natural-language question about the repository, which files should I open first?* In production workflows—triaging incidents, scoping changes, reviewing PRs, or steering an autonomous

agent—fast, correct file selection is often more valuable than generating a verbose explanation or speculative patch. The challenge is not building a one-off capable model; it is keeping that model faithful to a moving target. Even modest commit windows induce three distinct kinds of drift that systematically erode performance. **Structural drift** renames, moves, or deletes paths so that an answer that was correct yesterday becomes an alias or a tombstone today. **Surface churn** (typing, formatting, comments) inflates diffs without changing behavior, diluting signal for learning algorithms that treat all changes equally. **Behavioral drift** modifies logic within otherwise stable files, invalidating previously accurate associations between questions and files. Any update strategy that chases "freshness" risks catastrophic forgetting of reliable historical mappings; any strategy that prioritizes retention risks staleness on the most recent code. Our work is motivated by the need for a principled, repeatable way to maintain *both* freshness and retention—measured rigorously and achieved with predictable cost and latency.

We operationalize the task as **set-valued retrieval**. Given a natural-language question $qqq$, the model must produce a set of repository-root-relative file paths $\hat{Y} \subseteq P$, where $PPP$ is the finite universe of paths at the evaluation snapshot. This formulation is intentionally conservative: by constraining outputs to verifiable artifacts (exact, existing paths), we minimize hallucinations and make scoring unambiguous. Instead of open-ended text generation, the model performs closed-set, set-valued prediction. Two metrics capture the complementary goals: **Exact Match (EM)** requires the predicted set to equal the gold set; **Micro-averaged Recall (MR)** rewards partial recovery on multi-file answers and exposes under- or over-selection behavior. The central engineering question becomes temporal: *how do we maintain high EM and MR on new, old, and mixed test slices as the repository evolves?*

To answer this, we compare three families of update strategies that teams can deploy on a cadence:

**(1) Full Refresh @ HEAD.** Train a new repository-specific model on the latest snapshot. This yields a clean target and the best attainable "ceiling" when drift is large, but it incurs the highest training cost and wall-clock latency and is ill-suited to frequent, small updates.

**(2) In-Context Learning (ICL).** Keep model weights fixed and inject changes at inference time. We study two realizations: *diff-only* prompting (raw per-file git hunks) and *diff→English* prompting (compact, three-to-five sentence English summaries). As the commit distance widens and diffs grow noisy, we consistently observe English summaries outperform raw diffs by providing higher signal-to-noise in the prompt at the expense of prompt tokens and context-window pressure.

**(3) Incremental Fine-Tuning (Inc-FT).** Continue training on a **delta-derived dataset**— examples synthesized from files marked Modified/Added (M/A) in the commit window—while mixing a carefully controlled sample of OLD data to resist forgetting. We show that both the

*NEW:OLD ratio* and the *training schedule* (learning rate, epochs) are make-or-break: too little OLD and the model forgets, too much OLD and it underfits the delta.

A key contribution of this paper is an **evaluation and data-building protocol that respects structural drift** rather than punishing the model for remembering yesterday's names. We build an **alias map** from git metadata that records path renames and deletions (e.g., old/path.py → new/path.py or → __DELETED__). During scoring, we first remap predictions through this alias table so that a prediction of the *old* name for a *renamed* file receives credit at the *current* snapshot, while predictions that point to truly deleted files never receive credit. To probe residual "old memory" explicitly, we also introduce a **Forgetting Probe**: a diagnostic test set where every gold label underwent a structural change, evaluated *without* remapping, which quantifies raw emissions of old names. This alias-aware lens lets us separate **behavioral learning** (did the model internalize changed logic?) from **structural housekeeping** (did names merely move?), preventing misleading regressions when repositories reorganize.

Our empirical study spans **Flask, SQLAlchemy, Pandas, and Poetry**, and it proceeds in two phases. First, across repositories we establish the trade-offs between Full Refresh, ICL (diff vs. English), and Inc-FT under mixed, new-only, and old-only evaluations. Second, we conduct **targeted case studies** that answer practical questions engineers ask when choosing an update path. In **Poetry**, we *introduce and validate the alias-aware protocol alongside an index of changed files*—using the alias map during both **dataset construction** (to avoid rewarding deleted paths and to remap labels for renames) and **evaluation** (to score fairly at HEAD). Poetry is also where we run a head-to-head of **Inc-FT built from git-diff summaries versus Inc-FT built from full-file content**, revealing when surgical deltas win (rename/delete-heavy windows) and when broader file context is necessary (behavioral changes that diffuse across a file). In **Flask**, we perform an A/B effectiveness study that contrasts **Inc-FT on a (200→100)** **delta** with a **fresh Full Refresh at 100**, quantifying the residual "gap to ceiling" when teams opt for the cheaper, incremental path. Together, these studies provide complementary evidence: Poetry demonstrates how **alias-aware data and scoring** change conclusions in drift-heavy windows *and* how the choice of **git-diff vs. full-file Inc-FT** interacts with window character; Flask quantifies **how close Inc-FT can get to a full refresh** when behavioral changes dominate but structural churn is moderate.

Across the full suite of experiments, several patterns are consistent and actionable. On **mixed and new-only** sets, **Inc-FT with an OLD-aware mix** (e.g., NEW:OLD = 1:2 in Pandas 100→60, or modest pure-NEW in Flask's 200→100) often **matches or beats** ICL while yielding **lower runtime latency and token cost** (no long prompts). When training is infeasible or SLAs are tight, **ICL with English delta summaries** provides **immediate, controllable freshness**, particularly at larger commit distances, with the caveat of **prompt cost and context-window pressure**. **Full Refresh @ HEAD** remains the **ceiling** when sizable behavioral change accumulates; our Flask study quantifies the gap specifically so teams can decide whether the incremental path is "close enough" for their tolerance. Most importantly, the **Poetry alias-aware study** shows that, in rename/delete-heavy windows, naïve scoring can under- or over-state progress; **alias-aware construction and evaluation are necessary to draw the right conclusions** about both learning and forgetting.

The remainder of the paper proceeds as a self-contained narrative. We first ground the problem in the literature on evolving repositories, semantic code retrieval, in-context patching, and continual fine-tuning, highlighting where existing approaches struggle with structural drift and catastrophic forgetting. We then detail our **data and pipeline**—including base snapshot selection, delta construction, **alias map and index creation**, diff-to-English summarization, delta QA synthesis, and the **ICL and Inc-FT** mechanisms—so that the procedure is fully replicable. The **experiments** section presents results repository by repository, with **explicit subsections for the Poetry alias/index study** and the **git-diff vs. full-file Inc-FT** comparison, followed by the cross-repository tables for Flask, SQLAlchemy, and Pandas, and the **Flask effectiveness study** that frames Inc-FT's proximity to a full refresh. We close with an in-depth discussion of limitations (e.g., very large repos, window character detection, old-name emissions), ablation-backed recommendations for practice, and concrete future work.

# Related Work

The problem of finding "the right places to look" in a large, evolving repository sits at the intersection of classic code search, neural retrieval, repository-level program analysis, retrieval-augmented generation (RAG), and continual learning. Each tradition brings useful machinery, but each also leaves a gap that becomes painfully visible once the repo starts to move: indices go stale, embeddings drift, prompts overflow, and models forget. In this section we survey the lines of work we drew on most heavily, then situate our contributions—set-valued, path-only retrieval with time-aware data and scoring—against those threads. Throughout, we emphasize techniques that explicitly or implicitly address evolution: rename/delete churn, behavior changes that don't leave obvious lexical fingerprints, and the operational need to update systems on a cadence rather than in one heroic, annual retrain.

**Lexical and static code search.** The oldest family treats code as text and relies on inverted indices (Lucene/BM25) or lightweight static structure (e.g., symbol tables) to answer keyword queries over functions, classes, and files. These systems remain attractive because they are simple to build, cheap to update, and fast at query time. They struggle, however, on two fronts that matter in practice. First, *semantic aliasing*: the words developers use in queries rarely match identifiers in code, and the mapping is many-to-many ("delete user" vs. `remove_account`, `deactivate`, `archive`). Second, *set reasoning*: answers often span two or three files whose names share no terms (e.g., a request router and a serializer), yet lexical retrieval ranks files independently. Evolution compounds both problems. Every rename or module re-layout breaks string-level priors, and superficial edit churn (typing/formatting) perturbs token statistics without changing behavior. Lexical systems can be re-indexed at HEAD cheaply, but they provide no principled notion of "the *set* of files that, together, answer the question," and no protection against structural drift beyond re-crawling the tree.

**Neural code search and bi-modal encoders.** Learned encoders such as CodeBERT, GraphCodeBERT, UniXcoder, and their successors replace bag-of-words with joint embeddings of natural language and code. They shine at *instance-level* matching—"given this snippet or file, rank likely matches for this query"—and, importantly, they learn semantics that cross lexical gaps. Yet two tensions emerge for evolving repos. First, these models are usually optimized for *pointwise retrieval* of single snippets rather than *set-valued selection* over a finite, repository-specific label space. At deployment, practitioners typically retrieve the top-k files independently and then post-hoc group or filter, which is a weak surrogate for "pick the *minimal* set of files that jointly satisfy the information need." Second, temporal awareness is limited: embeddings baked at snapshot $t$ must be refreshed to reflect snapshot $t+\Delta$; otherwise, the ranker happily recommends now-deleted paths or misses a refactor that moved functionality. Teams respond by periodic re-embedding and re-indexing, but that treats evolution as noise rather than signal and offers no guarantees about catastrophic forgetting of older, still-relevant mappings.

**Repository-level graphs and structured retrieval.** A complementary line constructs richer, repository-level structures—call graphs, data-flow graphs, import graphs, or knowledge graphs of entities and relations—and retrieves *subgraphs* relevant to a query or a generation task. These methods are compelling because they encode *neighborhood priors*: if file A calls B and B constructs C, a query about A's behavior likely implicates B (and maybe C) even if names diverge. Graph-based systems power modern repository-level completion and review assistants by grounding an LLM with proximate code rather than raw text spans. Their Achilles heel under evolution is twofold. First, most graphs are *point-in-time* artifacts; keeping them fresh requires rebuilds that may lag HEAD, and graph deltas are not free when refactors touch wide swaths. Second, even when edges are updated, *scoring* typically ignores structural drift in labels: if `utils/json.py` was renamed to `json/provider.py`, a model that still emits the old path is penalized, even though it "remembered" the correct concept. Few systems make renames/deletes first-class citizens of the evaluation protocol.

**Prompting, RAG, and in-context updates.** With LLMs, the industry default became: keep a general model fixed and *inject freshness at inference* via retrieval. For code, this means selecting and chunking files, diffs, or documentation, then asking the model to answer or edit based on that context. RAG is attractive operationally—there is no training loop—and, when the retrieval is surgical, it delivers immediate lift on the newest code. But three practical limitations appear at scale. First, *context budgets*: real deltas contain many low-signal changes that crowd out the few lines that matter; summarizing diffs into concise English helps, but the summarizer must itself be version-aware. Second, *decision granularity*: the model typically returns prose or patches; coercing it to emit *only* paths under a strict output contract is possible but unnatural for generic chat prompts. Third, *latency/cost variance*: larger deltas or weaker retrieval expand prompts and make performance spiky. RAG fixes freshness but externalizes the combinatorial set selection problem to a brittle prompt.

**Continual and parameter-efficient fine-tuning.** A different branch embraces *updating the model* rather than the prompt. Adapter-based methods (LoRA/QLoRA, prefix-/prompt-tuning) make incremental training cheap enough to run often. In mainstream NLP, continual learning focuses on avoiding catastrophic forgetting across tasks or domains by mixing old and new data and regularizing updates. In code, however, evolution carries distinctive structure: labels (paths) literally *change names*; "new" behavior may live in files whose text barely changed; and much of the diff churn is non-behavioral. Prior work rarely teases apart *behavioral* vs *structural* change in either the *data* used for updates or the *metrics* used for scoring. As a result, teams either over-mix old data (preserving legacy at the expense of freshness) or over-fit to deltas (lifting new cases while silently erasing useful old mappings), without a principled read on what was gained or lost.

**Version-aware retrieval and refactoring-aware tooling.** A smaller, practice-driven literature acknowledges renames and deletes explicitly—e.g., mining `git log --name-status` to maintain alias tables, or building versioned indices keyed by file identity rather than path string. These systems handle lookups ("where did `foo.py` go?") and provenance ("which commits touched this concept?") well, but they stop short of *learning* a query→set-of-paths mapping and, crucially, they seldom integrate aliasing into *evaluation*. That omission matters: if the label space itself evolves, a fair scorer must credit a model that predicts an old name for a renamed file and must never reward deleted paths. Without alias-aware scoring, it is impossible to tell whether a drop in accuracy reflects lost *behavioral* knowledge or mere *string drift*.

---

**How our work differs.** Our formulation and pipeline borrow strengths from each tradition but close the gaps that hurt most in evolving repos. First, we *constrain the output space* to the finite, verifiable set of repository-root-relative paths at evaluation time and require the model to emit a *set* of paths—no prose, no snippets. This reframes the problem from open-ended generation or ranking to *set-valued retrieval*, making success (Exact Match, Micro-averaged Recall) unambiguous and directly useful to downstream tools that will open files. Second, we make *time* a first-class axis in both data and metrics. On the data side, our delta-derived examples are synthesized from files actually *Modified/Added* in the commit window, with OLD examples mixed under explicit ratios to control forgetting; when diffs are low-signal, we switch from line-diff supervision to *full-file* supervision to capture behavioral changes that the textual diff obscures. On the metrics side, we score at the target snapshot $Y$ using an *alias map* built from git metadata that maps old/path.py → new/path.py | "__DELETED__". Predictions are remapped through this table before EM/MR are computed, so a remembered rename is *credited* and deleted files are *never* rewarded; we also run a *Forgetting Probe* that disables remapping to quantify residual emissions of old names. Third, rather than canonizing a single update recipe, we treat *Full Refresh, ICL, and Incremental FT* as complementary levers with different operational trade-offs, and we provide head-to-head evidence for when each wins. In particular, we show (i) ICL with *English* delta summaries is the fastest path to freshness when training is infeasible, (ii) Inc-FT with *old-aware* mixes offers the best steady-state balance of new accuracy and retention, and (iii) *full-file* Inc-FT dominates *git-diff* Inc-FT on windows dominated by behavioral change, whereas the reverse holds in rename/delete-heavy windows.

Finally—and this is where our *Poetry* study is central—we integrate *version awareness into both training and scoring*. Poetry's 1739→1639 window featured heavy renames and deletions, making it an acid test for structural drift. We introduced the alias map and an index of M/A files there first, used them to curate NEW examples that *never* reward deleted paths and *always* remap old names to their Y-side equivalents, and scored all variants (ICL, Inc-FT: git-diff vs full-file, and Base@Y) with alias-aware metrics. That study demonstrates that once structural change is treated properly, the choice between git-diff and full-file supervision becomes legible: concise diff summaries excel when renames/deletes dominate; full-file supervision wins when behavior changes within otherwise stable files. This version-aware discipline then carries into our broader comparisons on Flask, SQLAlchemy, and Pandas, giving practitioners a playbook that acknowledges how code *actually* evolves rather than assuming a static world.

# Methodology

## Problem setting and drift

We study repository-aware **set-valued retrieval**. At time $t$, given a natural-language question $q$ about a codebase and a chosen target snapshot $Y$ (typically HEAD), the system returns a finite set of root-relative file paths $\hat{A} \subseteq P_Y$ that are jointly sufficient for a developer to investigate the question. Constraining outputs to **verifiable artifacts**—paths that actually exist at $Y$—makes evaluation crisp and forces the model to "speak the current dialect" of the repo.

Evolving codebases introduce three distinct forms of drift. **Structural drift** changes the label space itself through renames, moves, and deletions; **superficial churn** alters formatting, types, or comments with little behavioral impact; and **behavioral change** modifies logic while filenames and coarse structure remain stable. Our objective is twofold: (i) maximize freshness on questions anchored in changes between a base snapshot $X$ and the target $Y$; and (ii) preserve retention on questions about stable parts of the repo, all under realistic budget/latency constraints.

---

## Data & pipeline overview

We organize every experimental or production update as a **window** X⬚→⬚YX\!\to\!YX→Y. From this window we derive all artifacts needed by the three strategy families we evaluate—**Full Refresh @ HEAD**, **In-Context Learning (ICL)**, and **Incremental Fine-Tuning (Inc-FT)**—as well as the structures for **version-aware scoring**.

The pipeline proceeds as follows. We first materialize a **base** at XXX (including a base Q→paths dataset when needed) and compute a **delta manifest** from X⬚→⬚YX\!\to\!YX→Y, classifying files as Modified (M), Added (A), Renamed (R), or Deleted (D). From the manifest we build an **alias map** that resolves historical paths to their current identities (or to `__DELETED__`). In parallel we produce two delta representations: (a) raw unified diffs; and (b) compact **English delta summaries** purpose-built for models. These artifacts feed the three update strategies:

- **Full Refresh @ HEAD** trains a fresh adapter stack entirely on YYY.

- **ICL** freezes weights and injects the window's delta summaries at inference.

- **Inc-FT** continues training from the XXX base on a mixture of **NEW** examples derived from M/A files at YYY and carefully screened **OLD** examples from XXX that never touch those M/A files.

Evaluation is performed with **alias-aware EM/MR** at snapshot YYY, with an optional **Forgetting Probe** that disables remapping to quantify residual old-name emissions.

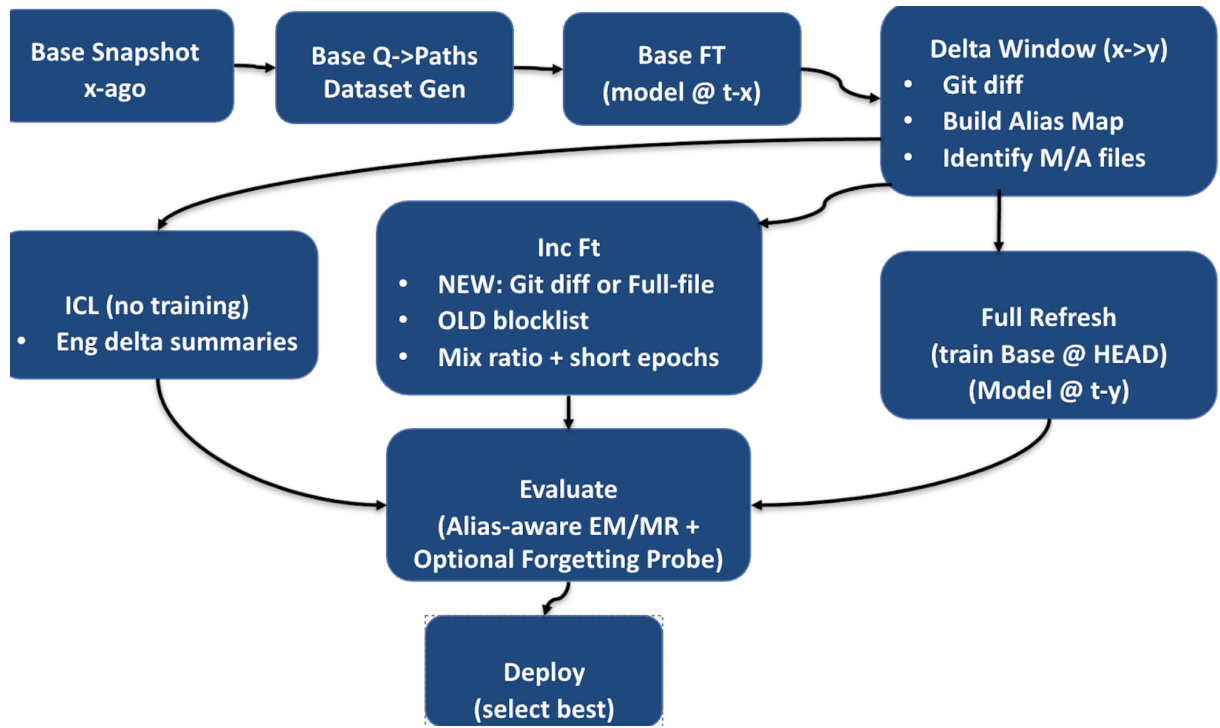

*Figure 1: — End-to-end data & pipeline*

# Version-aware index and alias map

Naïve string matching collapses behavioral forgetting with renames/deletes. We therefore materialize, **per window** X→YX→YX→Y, a machine-readable **change index** and an **alias map** that make structural drift first-class.

## Artifacts

Each window produces a single JSON with the following shape (only fields we actually use downstream are shown):

```
{

  "base": "<base_sha>",

  "head": "<head_sha>",
```

```json
  "changes":
[   {"status":"M","path":"examples/tutorial/flaskr/db.py","summary":"...",

{"status":"A","path":"tests/test_converters.py","summary":"..."},

{"status":"D","path":"flask/__init__.py","summary":"..."},
{"status":"R085","old_path":"flask/app.py","new_path":"src/flask/app.py","summary":"..."}
],

"alias_map": {

    "flask/__init__.py": "__DELETED__",

    "flask/app.py": "src/flask/app.py"

  },

  "adds":    ["…"],

  "mods":    ["…"],

  "deletes": ["…"],

  "renames":
[{"old":"flask/app.py","new":"src/flask/app.py","score":"R083"}]

}
```

- **changes[]** is one row per touched path (Git's `A|D|M|Rnnn` codes), with a **concise English summary** of the unified diff for that file.

- **alias_map** is the structural mapping we apply everywhere: old path → **new path** or → `"__DELETED__"`.

- The convenience lists (**adds/mods/deletes/renames**) drive dataset construction and guardrails (e.g., never label a deleted path).

**How we build it.**

We rely on `git` for the manifest and diffs, and a deterministic summarizer for English deltas. Rename detection is on; deletes are optional.

```
# Identify the repo root

git rev-parse --show-toplevel

# Window manifest with rename scores (e.g., R083) and code-only path
globs

git diff --name-status -M --diff-filter=ACMR  <X>..<Y> -- \

  "*.py" "*.ts" "*.tsx" "*.js" "*.jsx" "*.go" "*.rs" "*.java" "*.kt" \

  "*.c" "*.cc" "*.cpp" "*.h" "*.hpp" "*.php" "*.rb" "*.swift" "*.cs" \

  "*.sql" "*.sh" "*.bash" "*.zsh" "*.html" "*.css" "*.scss" >
/tmp/name_status.txt

# Per-file unified diffs (histogram, minimal, trim whitespace noise)

git diff -U5 --diff-algorithm=histogram --minimal --ignore-space-at-
eol -M <X>..<Y> -- <path>
```

Each diff is compressed to **3–5 sentences** with a reviewer-style prompt (temperature 0.2). We keep summaries terse, symbol-aware (mention functions/classes), and we **explicitly flag "formatting/no functional change"** when that's what the hunk shows. Long diffs are defensively truncated at MAX_DIFF_CHARS to bound latency.

**Summarizer system prompt (excerpt).**

```
You are a senior software engineer helping with code review.

You'll be given a unified git diff for a SINGLE file and its file
path(s).

Explain in 3–5 short sentences what the change does. Name affected
functions/classes/configs.

Avoid speculation; if unclear or formatting-only, say so.
```

The alias map credits remembered renames while never rewarding deleted paths, and the change summaries provide compact, high-signal supervision for ICL/Git-diff Inc-FT. Together they prevent misleading regressions in refactor-heavy windows and keep NEW/OLD mixing conflict-free.

---

# Delta construction: from Git diffs to model-ready signals

The NEW supervision signal must align with **meaningful change**, not line noise. For each M/A file in $X\{\cdot\}\to\{\cdot\}YX\!\to\!YX\to Y$, we extract the unified diff and compress it into a **3–5 sentence English summary** that names concrete symbols (functions, classes) and explicitly flags non-functional churn when present. These summaries are far more ICL-friendly as commit distance grows and also serve as anchors for synthesizing question→path examples for Inc-FT.

**Deterministic diff extraction:**

```
# Resolve the range once

BASE=HEAD~10

HEAD=HEAD

# Name-only list of code files changed in X..Y (rename detection on; A/C/M/R/D as needed)

git diff --name-only -M --diff-filter=ACMRD $BASE..$HEAD -- \

  "*.c" "*.cc" "*.cpp" "*.h" "*.hpp" "*.rs" "*.go" "*.py" "*.rb" "*.php" \

  "*.java" "*.kt" "*.scala" "*.cs" "*.m" "*.mm" "*.swift" \

  "*.js" "*.jsx" "*.ts" "*.tsx" "*.vue" \

  "*.sh" "*.bash" "*.zsh" "*.sql" \

  "*.html" "*.css" "*.scss" "*.sass" \
```

```
    > /tmp/changed_paths.txt

# For any path p in that list: unified diff, trimmed of whitespace
noise, rename-aware

git diff -U5 --diff-algorithm=histogram --minimal --ignore-space-at-
eol -M $BASE..$HEAD -- "$p" > /tmp/diff.patch
```

**Prompt (Diff → English summary, per file).**

```
System: You are a senior engineer doing code review for exactly ONE
file.
```

Given (a) its repository-relative path and (b) the unified git diff for window X→Y,

```
describe in 3-5 short sentences WHAT changed. Name impacted
functions/classes.

If the diff is mostly formatting, comments, or types, say so
explicitly.

Do not speculate beyond the diff; do not restate the lines.

Inputs:

- path: {repo_path}

- unified_diff: ```patch

{diff_text}
```

Output: 3–5 sentences, terse, in English.

```
**Output artifact.** The script writes a JSON like:

```json

{

  "examples/tutorial/flaskr/db.py": "Converts single quotes to double
quotes in ... no functional change.",

  "tests/test_converters.py": "Adds tests for custom URL
converters ...",

  "flask/__init__.py": "Deleted file; exports moved into
src/flask/__init__.py ..."

}
```

- We truncate very long diffs (configurable) before summarization to keep requests bounded.
- If a diff is effectively noise (formatting/comments), the summary says so; such deltas can be down-ranked by the retriever to protect the context budget.

For windows where behavior emerges from non-local context (e.g., a new guard whose implications are spread across methods), we also build **Full-file views** at YYY and drive question synthesis from entire files rather than line deltas.

---

# In-Context Learning (ICL) specifics

ICL keeps the base model frozen and **patches its knowledge at inference** using the window's delta summaries. A lightweight retriever selects the handful of most relevant summaries for a given question (the version-aware index prevents brittle filename matching). The resulting prompt concatenates: (i) system rules that enforce our output contract; (ii) the **Delta Updates** (English summaries); and (iii) the user's question. This strategy delivers **instant freshness** without training cost; its limits are prompt budget, latency, and the quality of summarization/selection.

**Prompt (ICL inference, English-delta variant).**

<|im_start|>system

You are a codebase assistant with prior knowledge of this repository (from fine-tuning).

At inference time, you are ALSO given *Delta Updates* that describe changes which occurred

after your training snapshot. Treat these updates as patches that override or extend your

existing knowledge wherever there is any conflict.

Your task: given a user question, return repository-root-relative file path(s) that are most relevant.

Strict output rules:

- Output ONLY a JSON array of unique strings with exact, complete, root-relative paths (no leading '/', './', '../').

- Do NOT include any commentary, reasoning, explanations, or trailing text. Example: ["a.py","dir/b.py"]

- If you are not confident any paths are relevant or still exist, return [].

Selection guidelines (use silently; do not output this text):

- Use your fine-tuned repo knowledge as the base. Apply the Delta Updates below as the source of truth for recent changes.

- Prefer current paths and behaviors described in the Delta (e.g., renames, file additions/removals, major refactors).

- It is valid to select files NOT mentioned in the Delta if they are relevant per your existing repo knowledge.

- Exclude files that the Delta indicates were deleted or replaced (unless the question is explicitly historical).

- Favor primary implementation files before tests/docs unless the question targets tests/docs.

```
- Be conservative: only return paths you are confident actually exist
in the current repo state.

Delta Updates (changes since your training snapshot):

<delta_info>

{delta_info}

</delta_info>

<|im_end|>

<|im_start|>user

Question: {question} /no_think

<|im_end|>

<|im_start|>assistant
```

We found **English summaries** to dominate **raw diffs** as windows widen: token-count stays bounded while essential semantics are preserved. Raw diffs remain useful for very small windows or when summaries are not yet available.

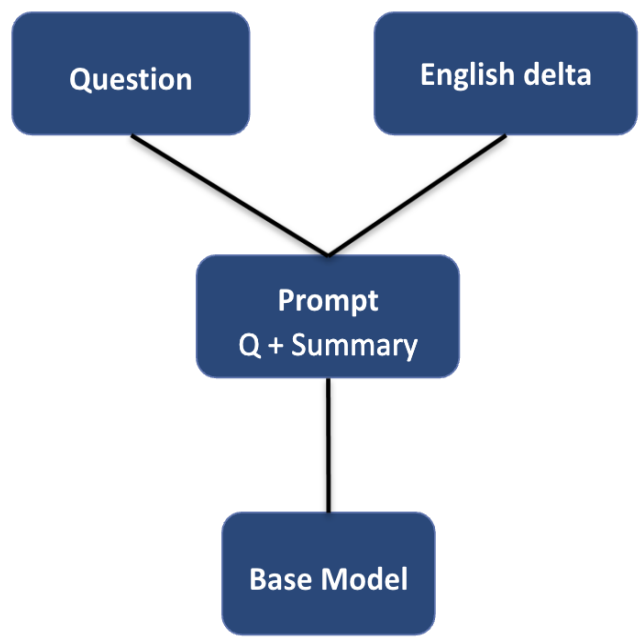



---

# Incremental Fine-Tuning (Inc-FT) specifics

Inc-FT continues training from the XXX base on a carefully curated dataset. The **NEW pool** contains only examples anchored on M/A artifacts at YYY; labels are first passed through the alias map to ensure they are **Y-valid**. The **OLD pool** is sampled from XXX with a strict **blocklist** that excludes any example touching changed files; this prevents teaching the model two inconsistent identities for the same concept. We train parameter-efficient adapters (LoRA/QLoRA) with **short schedules** and **mid/low learning rates**, varying the **NEW:OLD ratio** to navigate the freshness–retention frontier.

## Git-diff Inc-FT (surgical)

Here, question synthesis uses the **English diff summaries** as supervision anchors. Prompts bias toward naming the changed file explicitly and—when appropriate—include companions (tests, providers, serializers) to elicit multi-path answers.

**Prompt (synthesize a NEW training pair from a diff summary).**

```
You are given:

- The repository-relative file path that changed (pre/post as
relevant)

- A concise English summary describing how that file changed (derived
from a git diff)

Your task:

- Propose up to {target} specific developer questions about the change
in this file.

- Each question must include at least one changed symbol
(function/class/param/constant) in backticks.

- Each item must include a minimal list (1–{max_files_per_q}) of
repository-root–relative paths relevant to answering it.
```

```
- Output ONLY this JSON shape (no extra keys, no prose):

{{

  "samples": [

    {{

      "question": "Developer question here with at least one
`ChangedSymbol`",

      "relevant_file_paths": ["file1.py", "dir/file2.py"]

    }}

  ]

}}

Constraints:

- Focus strictly on the described changes (not the whole repo).

- Use repo-root-relative UNIX paths that actually exist.

- No file names or paths in the *question* text (symbols OK).
```

**When it shines.** Rename/delete-heavy windows or low-signal churn. The supervision targets the precise locus of change, minimizing interference with legacy behavior.

## Full-file Inc-FT (holistic)

Here, we ignore line-level diffs and condition question synthesis on the **entire file content at YYY**. This captures behaviors that span methods or rely on broader class invariants.

**Prompt (synthesize from full file).**

```
You are a senior software engineer analyzing a codebase.
```

Given:

1) The repository-root-relative path of the current file

2) The entire contents of the current file

Your task:

- Generate up to {max_per_file} realistic, high-quality developer questions.

- Each question should require understanding the current file (and other files when natural).

- For each question, output ONLY the minimal set of file paths (1–{max_files_per_q}) that are relevant.

- Paths MUST be repo-root-relative, use "/", exist, be sorted & unique.

Return ONLY:

```
{{
  "samples": [
    {{"question": "Developer question here", "relevant_file_paths": ["file1.ext"]}}
  ]
}}
```

(No prose.)

**When it shines.** Behavioral changes within files where line diffs are noisy or insufficiently descriptive (e.g., policy shifts spread across a class).

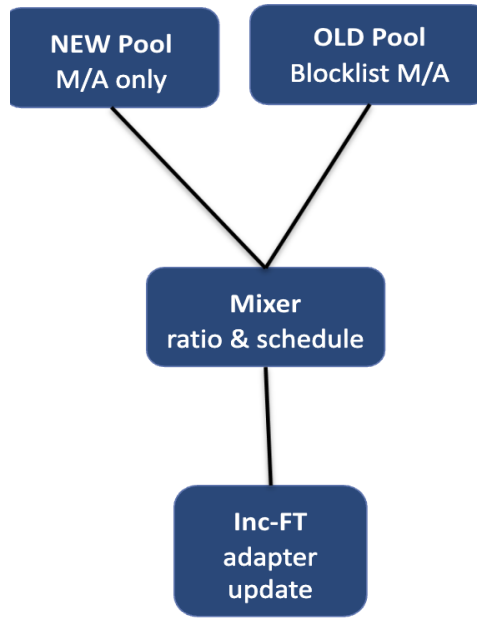

*Figure 3: Inc-FT composition*

# Full Refresh @ HEAD

Full Refresh trains a new adapter stack entirely on YYY using a comprehensive dataset that reflects the current label space. It achieves the cleanest **upper bound on NEW** because supervision and labels are perfectly aligned with YYY, avoids runtime prompt bloat, and simplifies operational reasoning. Its cost is **training wall-clock and cadence**. In practice, teams combine periodic Full Refreshes (e.g., after large refactors or major releases) with more frequent ICL/Inc-FT steps between them.

# Metrics and evaluation (alias-aware)

We evaluate the system at a target repository snapshot YYY (typically HEAD) on three disjoint slices:

- **NEW-only** — all gold paths were **Modified/Added** (M/A) in the window

  $X_{\square} \to _{\square} YX\backslash !\backslash to\backslash !YX \to Y.$

- **OLD-only** — all gold paths are unchanged across $X_{t} \to_{t'} Y$X\!\to\!YX→Y.

- **MIXED** — at least one gold path is NEW and at least one is OLD.

All outputs are **sets of root-relative paths that exist at $Y$YY**, and order is irrelevant. Throughout, let $P_Y$P\_YPY be the finite set of paths that exist at snapshot $Y$YY.

**Instance setup**

For each question $i$iii:

- Gold set $G_i \subseteq P_Y$Gi⊆PYG\_i \subseteq P\_YGi⊆PY — the minimal set of files a developer should open at $Y$YY.

- Model prediction $\hat{A}_i$A^i\hat{A}\_iA^i — a set of paths produced under the paper's output contract.

We collapse duplicates in $\hat{A}_i$A^i\hat{A}\_iA^i and drop non-string items before scoring.

---

**Exact Match (EM)**

**Definition.** EM is the fraction of questions whose **entire** predicted set matches the gold set exactly.

Per-instance indicator:

$EM_i = \mathbf{1}\{\,\hat{A}_i = G_i\,\}$EMi = 1{ A^i=Gi }.\text{EM}\_i \;=\; \mathbf{1}\{\,\hat{A}\_i = G\_i\,\}.EMi=1{A^i=Gi}.

Corpus-level EM:

$EM = \frac{1}{N}\sum_{i=1}^{N} \text{EM}_i$EM = 1N∑i=1NEMi.\text{EM} \;=\; \frac{1}{N}\sum\_{i=1}^{N} \text{EM}\_i.EM=N1i=1∑NEMi.

**Interpretation.** EM is unforgiving but crisp: you get credit only if you return **all and only** the gold paths. It's the right measure when downstream tools will open exactly what you output.

---

**Micro-averaged Recall (MR)**

**Definition.** MR rewards **partial recovery** on multi-file answers by summing true positives over all instances and normalizing by the total number of gold paths.

Per-instance recall:

$$r_i = |A\hat{}inGi||Gi|.r\_i \;=\; \frac{|\hat{A}\_i \cap G\_i|}{|G\_i|}.ri=|Gi||A\hat{}inGi|.$$

Corpus-level micro recall:

$$MR = \sum i=1N|A\hat{}inGi|\sum i=1N|Gi| = \sum i|TPi|\sum i|Gi|.\text{MR} \;=\; \frac{\sum_{i=1}^{N} |\hat{A}\_i \cap G\_i|}{\sum_{i=1}^{N} |G\_i|} \;=\; \frac{\sum_i |TP\_i|}{\sum_i |G\_i|}.MR=\sum i=1N|Gi|\sum i=1N|A\hat{}inGi|=\sum i|Gi|\sum i|TPi|.$$

**Interpretation.** MR reflects how often a developer gets at least the essential file(s), even if the set isn't perfect. EM and MR together expose **under-selection** (good precision but missed files → EM↓, MR somewhere in between) versus **over-selection** (many extras → MR can be high while EM stays low).

|  | EM | MR |
|---|---|---|
| What | Exact file set | Recall across labels |
| Why | Strict correctness | Robustness to partial |

*Figure 4: Metrics (EM and MR)*

---

**Why alias-aware scoring is necessary**

Code evolves structurally: files can be **renamed/moved** (RRR) or **deleted** (DDD). A model that still "speaks the XXX dialect" may predict `flask/app.py` when the file now lives at `src/flask/app.py`. Penalizing that as wrong would conflate **behavioral forgetting** with **string drift**. Conversely, we must never reward predictions of truly **deleted** paths.

We therefore score predictions **through an alias map** built from the window $X\square\to\square YX\!\to\!!YX\to Y$.

**Alias-aware transformation and scoring**

Before computing EM/MR, we **normalize** each prediction:

1. **Normalize path strings.** Strip duplicates; enforce repository-root-relative.

2. **Remap via alias.** For each $p \in \hat{A}_i$ $p \in \hat{A}_i$ $p \in A^i$,

   $p' = \{alias(p)if p \in dom(alias),potherwise (assume already a Y-side path).$ $p' \;=\;$ $\begin{cases} \text{alias}(p) & \text{if } p \in \text{dom}(\text{alias})$, $\\ p & \text{otherwise (assume already a } \(Y\)\text{-side path)}\}. \end{cases}$ $p'=alias(p)if p \in dom(alias),otherwise (assume already a Y-side path).$

3. **Drop deletions.** Remove any $p'$ $p'$ that equals `__DELETED__`.

4. **Clamp to PYP_YPY.** Discard any remaining $p' \notin PY$ $p'$ $\notin P\_Y$ $p' \in /PY$.

Let the remapped set be $A^i$ $\hat{A}_i$ $A^i'$. We then compute EM and MR **on $A^i$ $\hat{A}_i$ $A^i'$ vs $G_i$ $G_i$ $G_i$** using the formulas above.

**Worked example.**
Gold at YYY: $G=\{src/flask/app.py\}$ $G = \{\texttt{src/flask/app.py}\}$ $G=\{src/flask/app.py\}$.
Model outputs $\hat{A}=\{flask/app.py,README.md\}$ $\hat{A}=\{\texttt{flask/app.py}, \texttt{README.md}\}$ $A^=\{flask/app.py,README.md\}$.
Alias has flask/app.py → src/flask/app.py.
After remap + clamp:
$A^'=\{src/flask/app.py\}$ $\hat{A}'=\{\texttt{src/flask/app.py}\}$ $A^'=\{src/flask/app.py\}$ (assuming `README.md` isn't in gold).
Scores: EM $=1=1$, MR $=1=1$. The model remembered the right **concept** under the old name; we credit it.

If instead the model predicted a deleted file: flask/__init__.py → __DELETED__, it is **dropped** before scoring (no credit, no penalty beyond not helping EM/MR).

**Forgetting Probe (no-remap diagnostic)**

To measure **residual structural memory** explicitly, we construct a diagnostic slice where **every** gold label underwent a structural change (rename or delete). We then **disable alias remapping** and score against **old** names:

- **Old-name EM/MR:** apply the same formulas with gold sets expressed in XXX-side paths and **no** alias.

- **Old-path emission rate:** the share of predicted paths that are in the set D = \{p \mid \text{alias}(p)=\texttt{__DELETED__}\} or in dom(alias)\PY\text{dom(alias)}\setminus P_Ydom(alias)\PY.

This probe answers: *"How often does the model still speak the XXX dialect when structure moved?"* High emission rates motivate more OLD-aware mixing or a brief full refresh.

**What alias reasons tell us**

Alongside EM/MR, we tabulate how hits were achieved:

- `direct` — prediction was already in PYP_YPY.

- `alias_rename` — hit arrived via rename/move.

- `alias_deleted` — predicted a deleted path (dropped; counted only in diagnostics).

- `rescued_*` — conservative heuristic rescues, used for **analysis only** (not credited in scoring).

These breakdowns separate **behavioral learning** (direct hits) from **structural housekeeping** (alias rescues), preventing misleading conclusions in rename/delete-heavy windows.

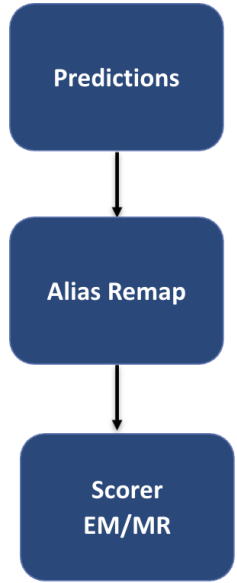

*Figure 5: Alias-aware evaluation*

# Experimental Results

This section evaluates the three update strategies introduced earlier—**Full Refresh**, **In-Context Learning (ICL)** with *raw diffs* or *English delta summaries*, and **Incremental Fine-Tuning (Inc-FT)** built from delta-derived datasets—across multiple open-source repositories. All experiments share a single task and output contract: given a natural-language question about a repository at state YYY (e.g., HEAD), the model must emit a JSON set of root-relative file paths that exist at YYY. We report **Exact Match (EM)** and **Micro-averaged Recall (MR)** and, unless otherwise stated, apply the **alias-aware scorer** to credit renames while never rewarding deleted paths. For clarity, we present per-repository results first, followed by cross-cutting analyses and practical guidance.

## Flask

**Setup.** We fine-tuned a base model on a snapshot approximately 10 commits behind HEAD, synthesized delta examples to the current state, and evaluated three splits: **48 MIXED** (24 old + 24 new), **24 NEW-only**, and **48 OLD-only**. For Inc-FT we swept several NEW:OLD ratios and kept schedules short to control forgetting.

## Mixed (24 old + 24 new)

| Variant | EM | MR | ΔEM vs Base | ΔMR vs Base |
|---|---|---|---|---|
| Base (10-ago) | 0.5417 | 0.5185 | — | — |
| ICL: diff-only | 0.6667 | 0.6296 | +0.1250 | +0.1111 |
| ICL: diff→English | 0.6250 | 0.6481 | +0.0833 | +0.1296 |
| Inc-FT 96n/144o | 0.6458 | 0.6296 | +0.1041 | +0.1111 |
| **Inc-FT 96n/192o** | **0.7500** | **0.7407** | **+0.2083** | **+0.2222** |
| Inc-FT 96n/96o | 0.2708 | 0.2407 | −0.2709 | −0.2778 |

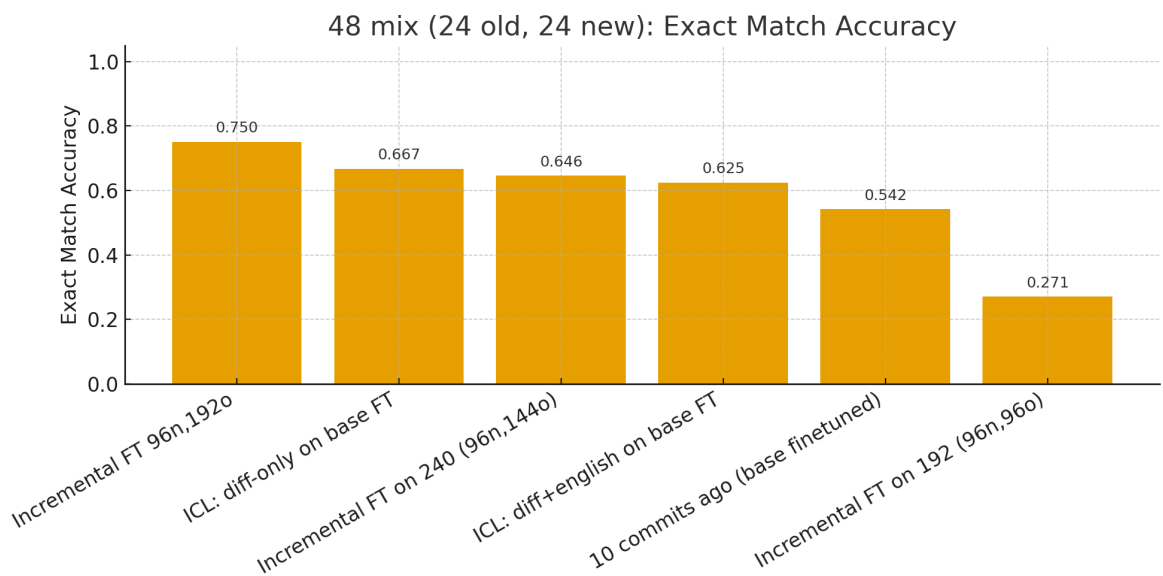

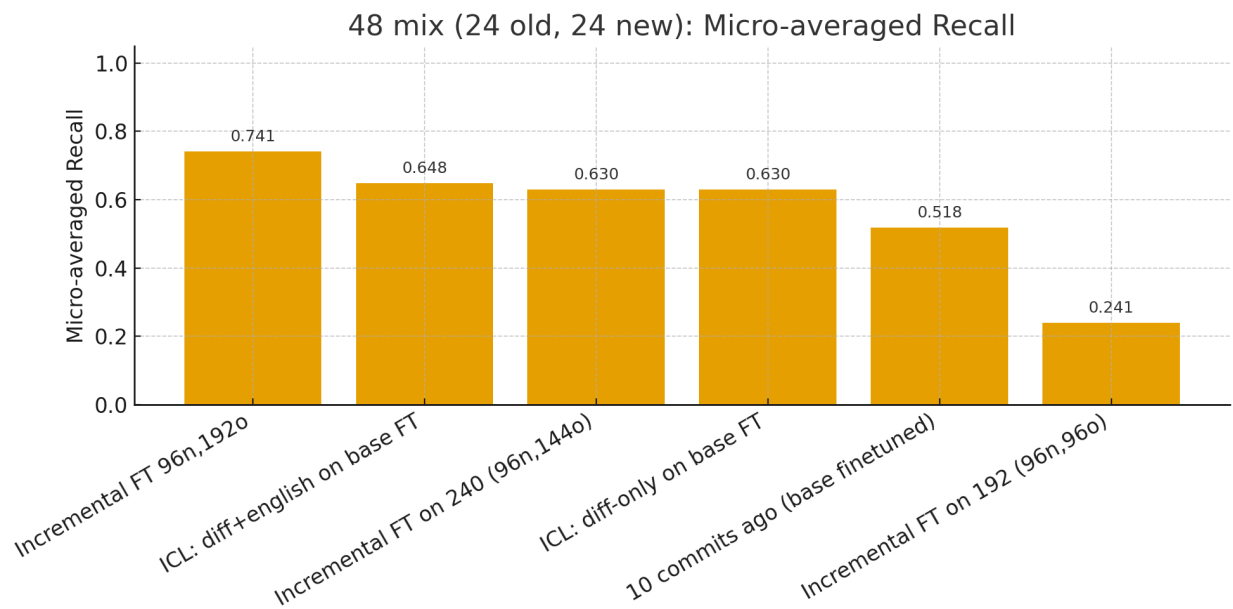

48 mix (24 old, 24 new): Micro-averaged Recall

**Observation.** Inc-FT with a modestly **old-leaning mix (96n/192o)** dominates both EM and MR on the mixed set, suggesting that a small amount of curated OLD stabilizes behavior while NEW anchors the delta.

## New-only (24)

| Variant | EM | MR | ΔEM | ΔMR |
|---|---|---|---|---|
| Base (10-ago) | 0.5417 | 0.5417 | — | — |
| ICL: diff-only | 0.8750 | 0.8750 | +0.3333 | +0.3333 |
| ICL: diff→English | 0.7917 | 0.8750 | +0.2500 | +0.3333 |
| Inc-FT 96n/144o | 0.8333 | 0.8333 | +0.2916 | +0.2916 |
| **Inc-FT 96n/192o** | **0.9583** | **0.9583** | **+0.4166** | **+0.4166** |

| Inc-FT 96n/96o | 0.2917 | 0.2917 | −0.2500 | −0.2500 |
|---|---|---|---|---|

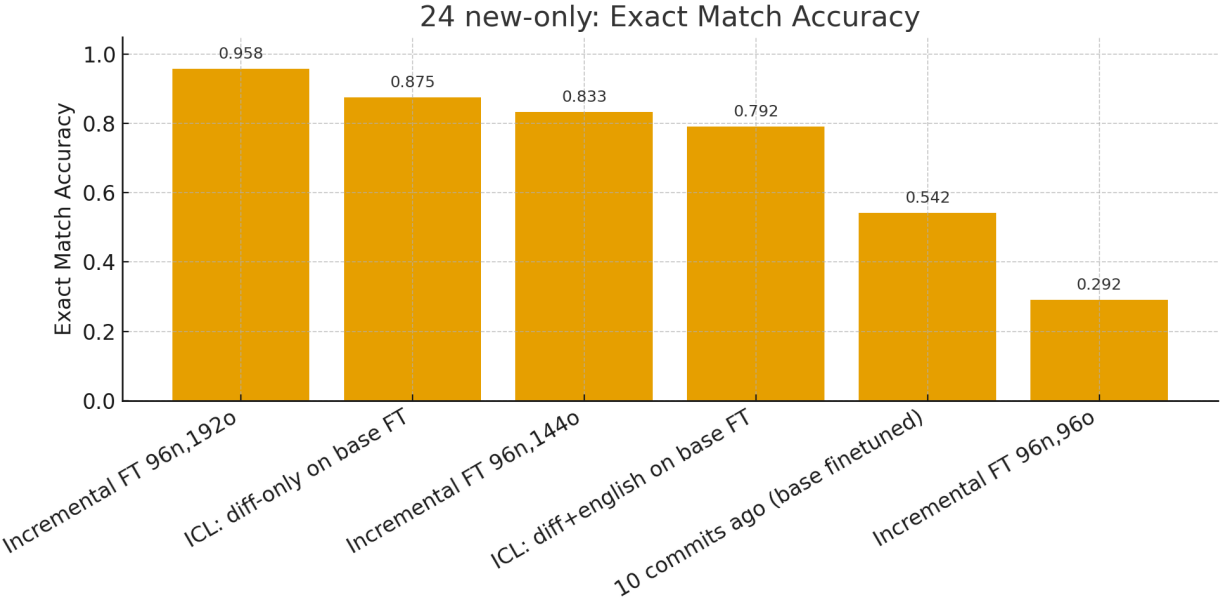

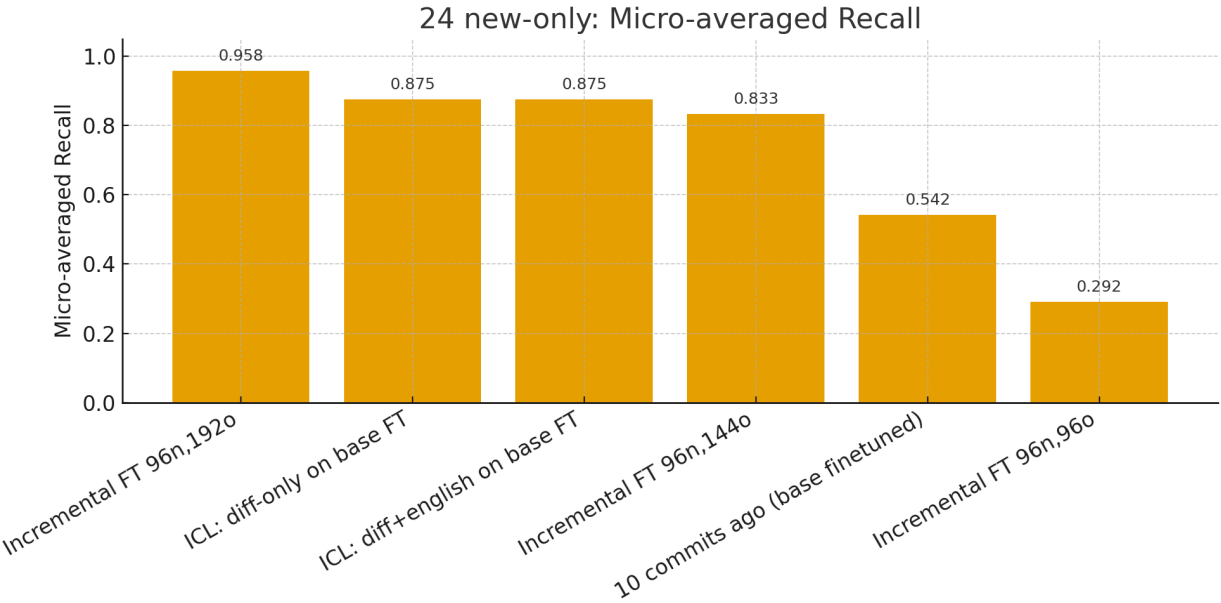

**Observation.** For **NEW-only**, Inc-FT (96n/192o) is best, with **ICL** offering a strong no-training alternative (especially diff-only in this window).

## Old-only (48)

| Variant | EM | MR | ΔEM | ΔMR |
|---|---|---|---|---|
| Base (10-ago) | 0.4792 | 0.5254 | — | — |
| Inc-FT 96o/96n | 0.1250 | 0.1017 | −0.3542 | −0.4237 |
| **Inc-FT 144o/96n** | **0.4792** | **0.5593** | **+0.0000** | **+0.0339** |
| Inc-FT 96n/192o | 0.4792 | 0.5085 | +0.0000 | −0.0169 |

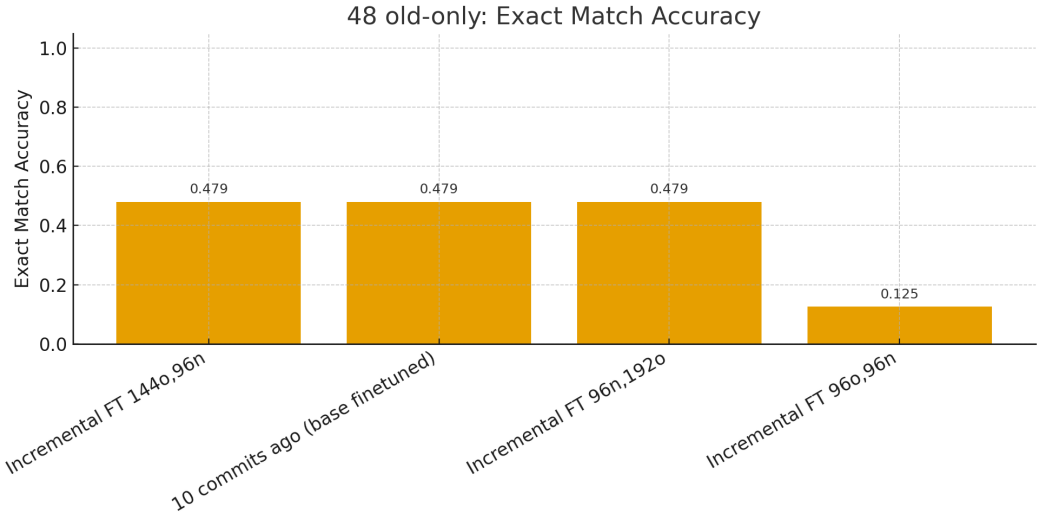

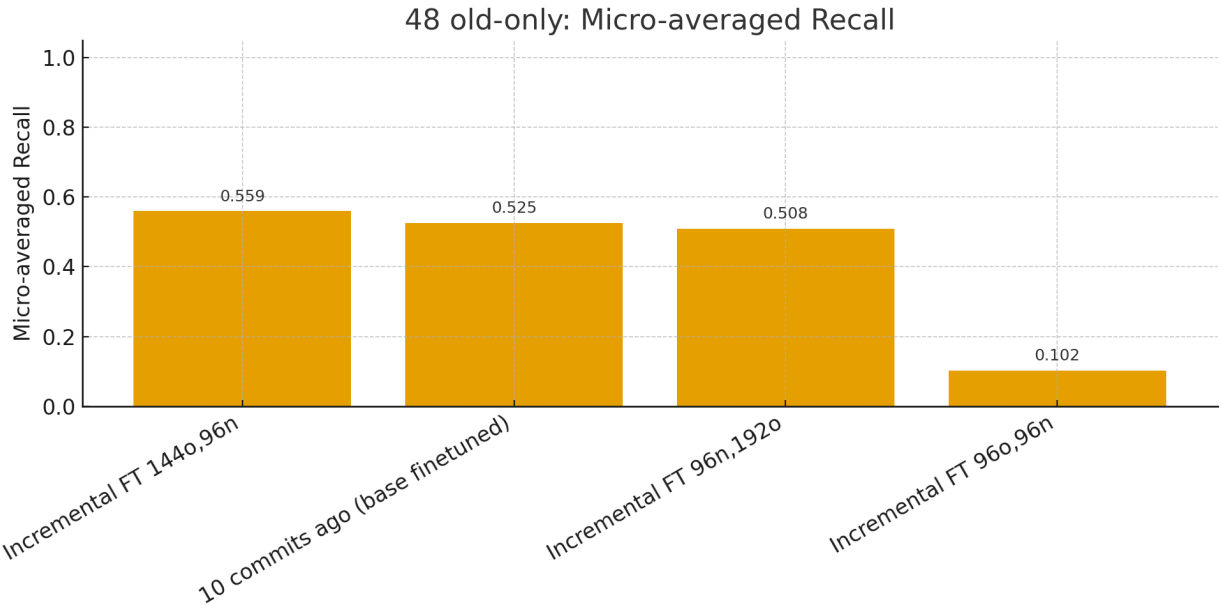

**Observation. OLD retention** is maximized by an **OLD-aware** mix (144o/96n), achieving MR gains at EM parity with the Base.

**Takeaways (Flask).**
 (i) **Best overall** on MIXED/NEW: **Inc-FT 96n/192o**.
 (ii) **Fastest freshness** with **no training**: **ICL (diff-only)**.
 (iii) Avoid **96n/96o**: consistent underfitting and forgetting.

---

# SQLAlchemy

**Setup.** We rolled the base back by ~50 commits and evaluated deltas to several targets (e.g., 50→40, 50→30, 50→20, 50→10). We compared **Inc-FT** against **ICL (raw diffs vs English summaries)** on MIXED/NEW/OLD splits. Below we include the dashboard visuals and summarize the quantitative takeaways:

**High-level observations:**

- On **MIXED**, **Inc-FT** generally outperforms ICL; among ICL variants, **English summaries** surpass **raw diffs** as commit distance grows (cleaner signal, less prompt noise).

- On **NEW-only**, **Inc-FT** and **ICL-English** both lift over the base; **ICL-raw** degrades as diffs lengthen.

- On **OLD-only**, **Inc-FT** retains better than ICL; raw diffs are brittle.

## Dashboard

### Inc-FT vs ICL (English vs Raw Diff) on Mixed se

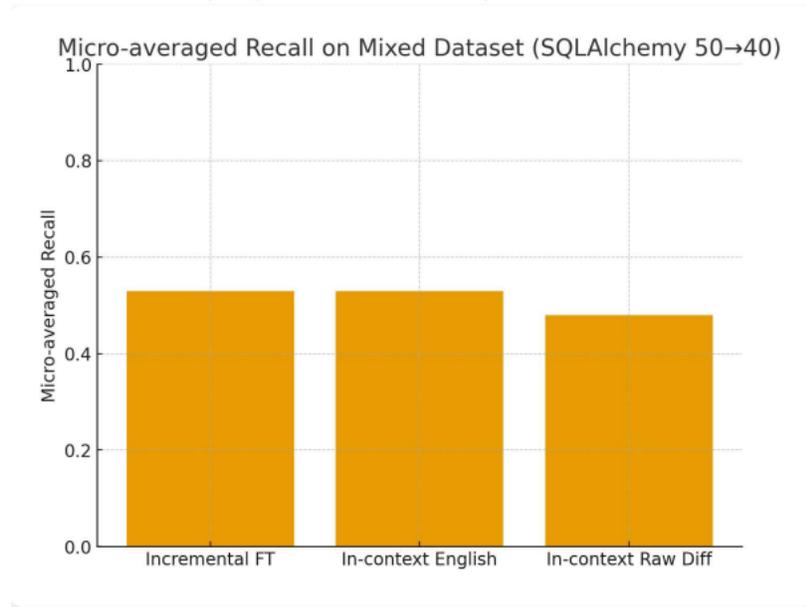

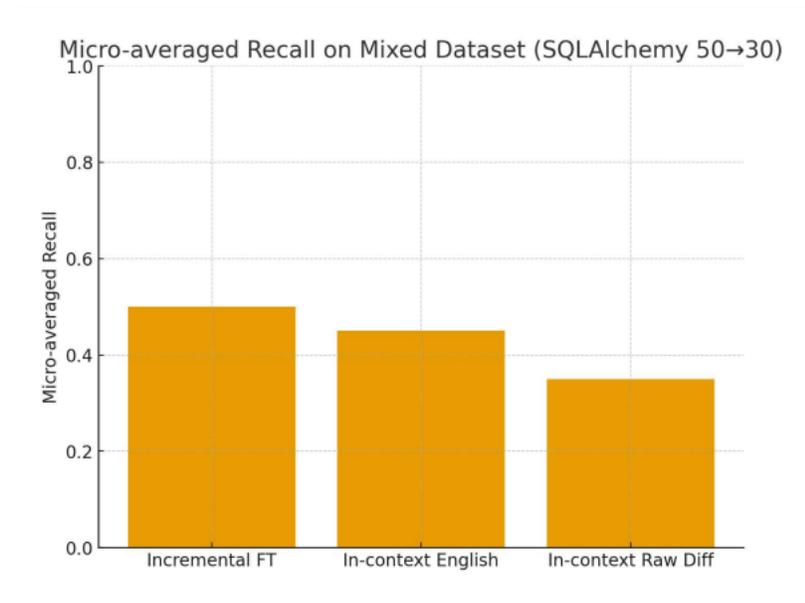

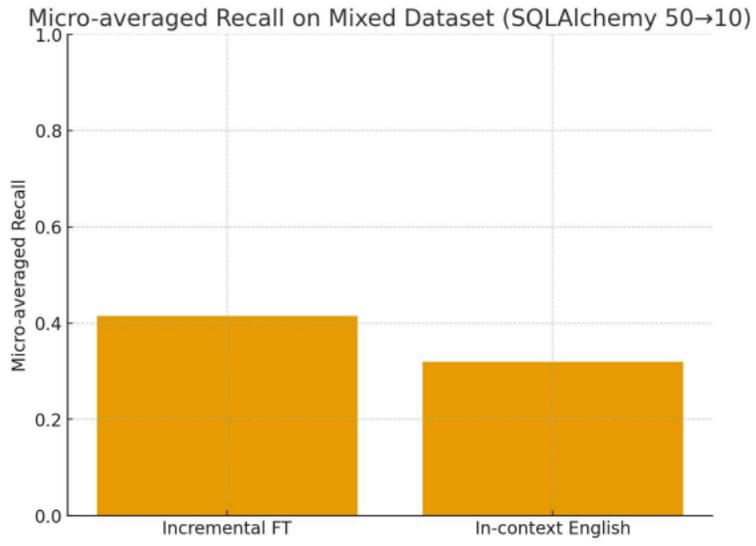

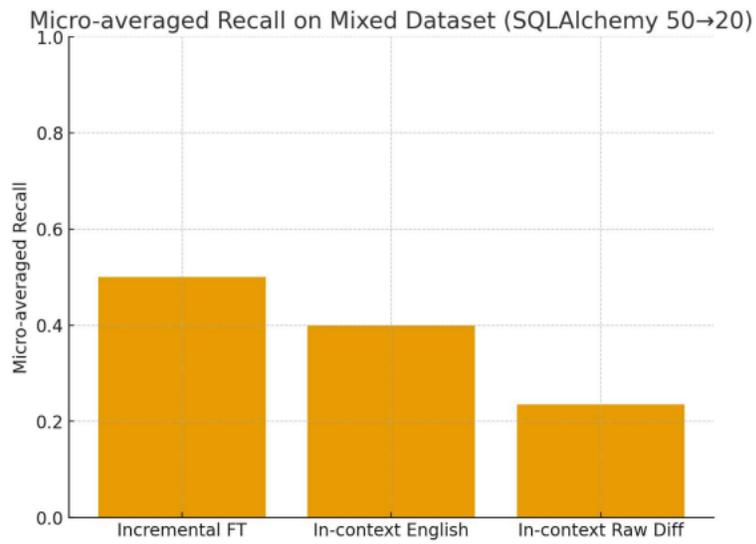

**Strategy performance across dataset types**

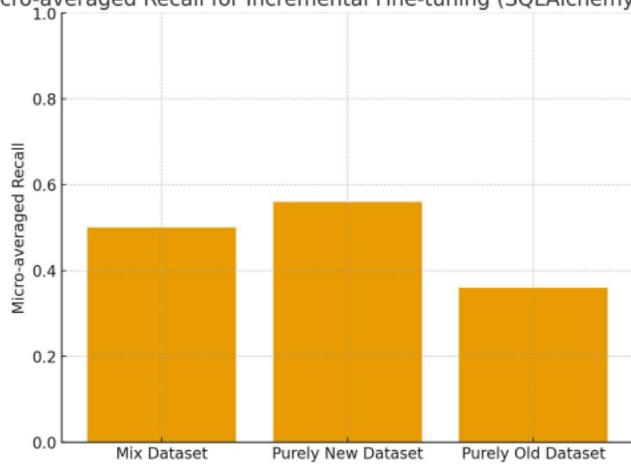

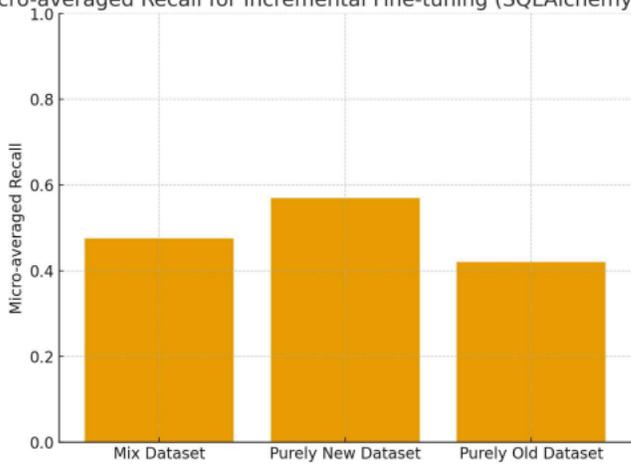

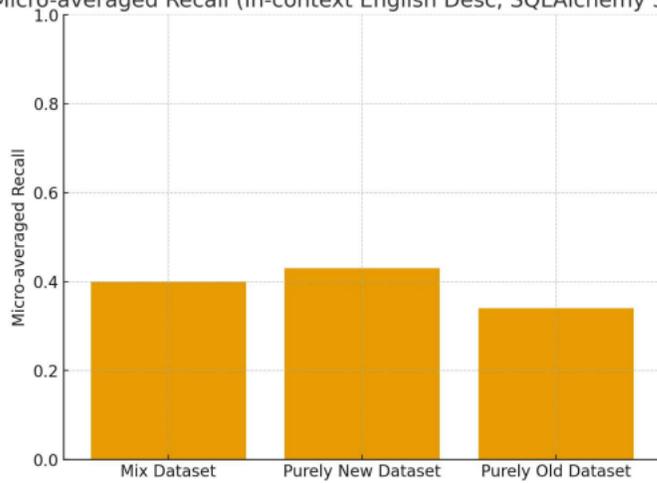

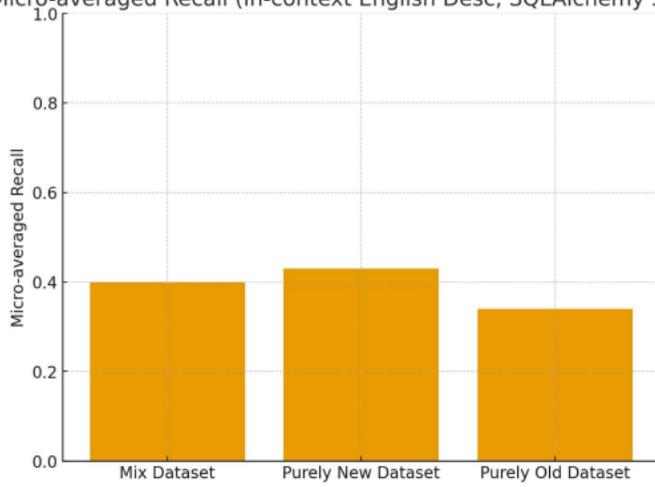

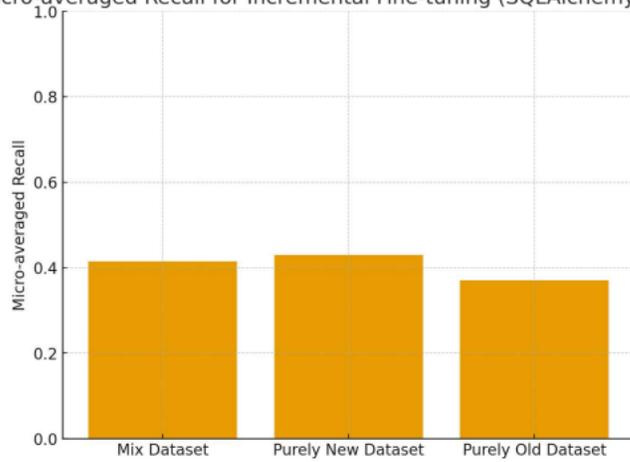

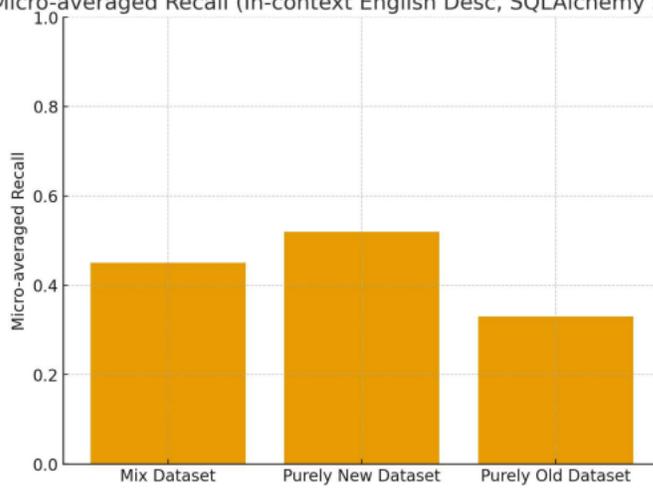

**Micro-averaged Recall across model/test variants**

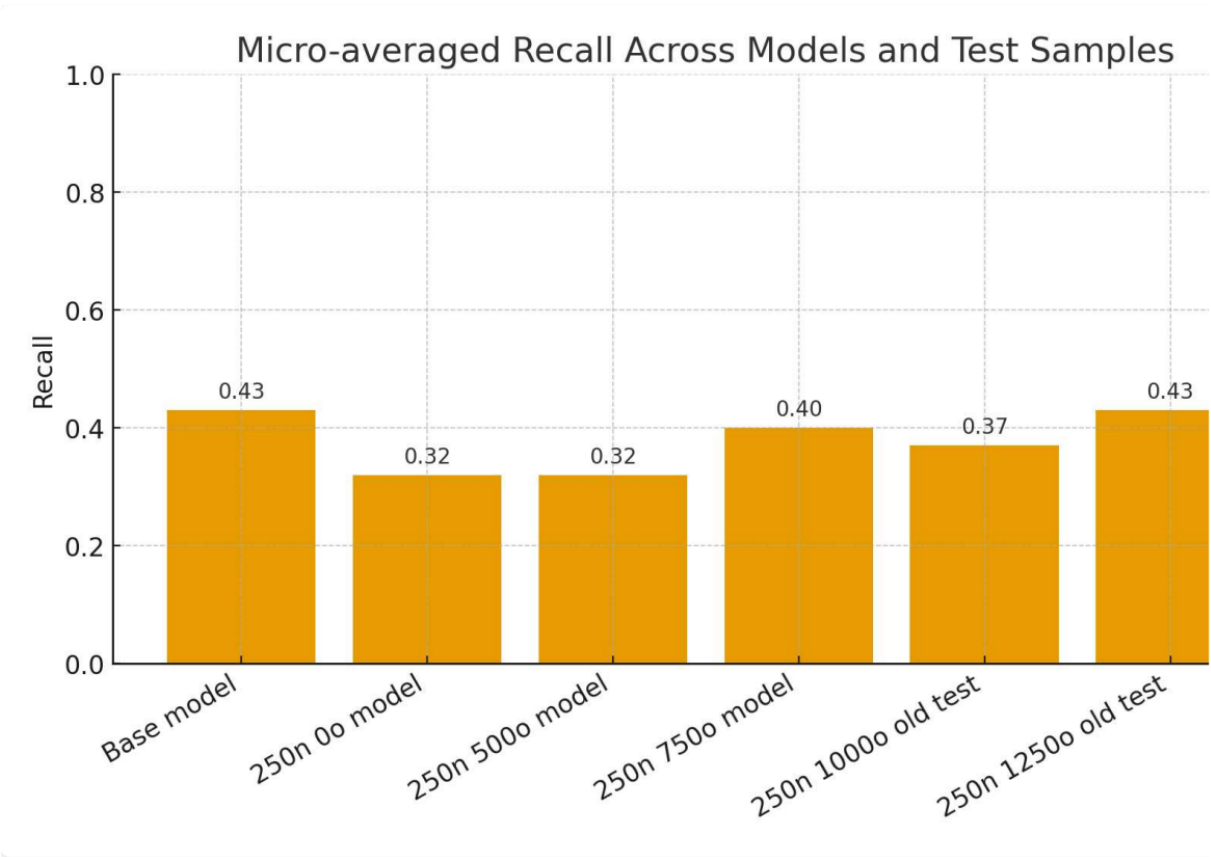

**Other graphs: MR and EM by dataset type**

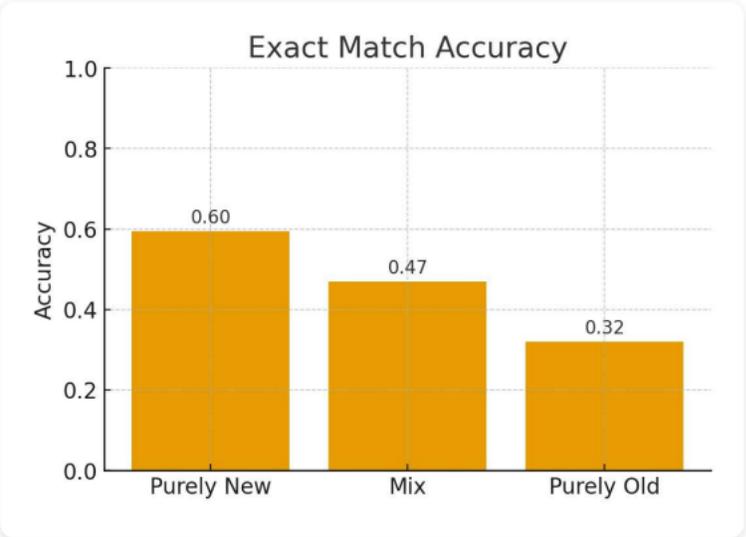

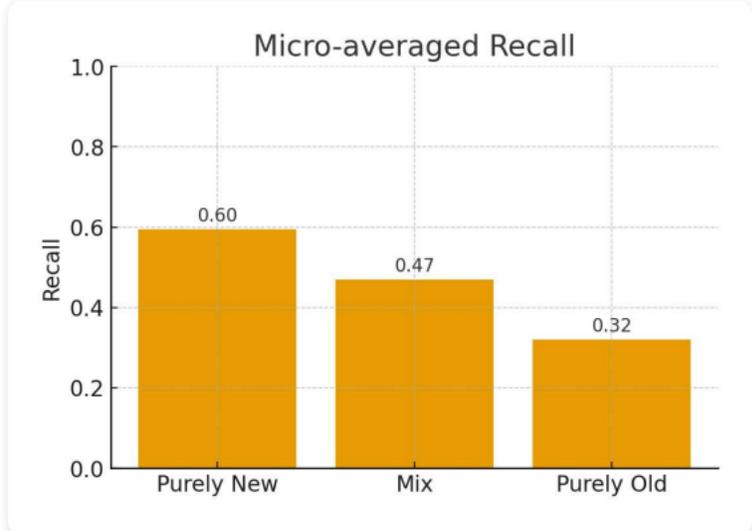

**Exact Match Accuracy — new-only vs old-only**

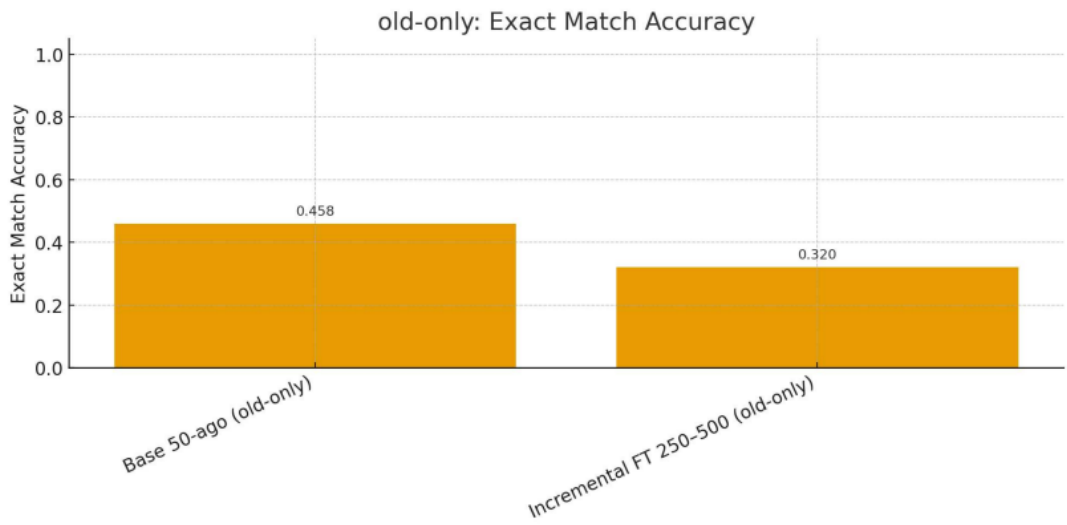

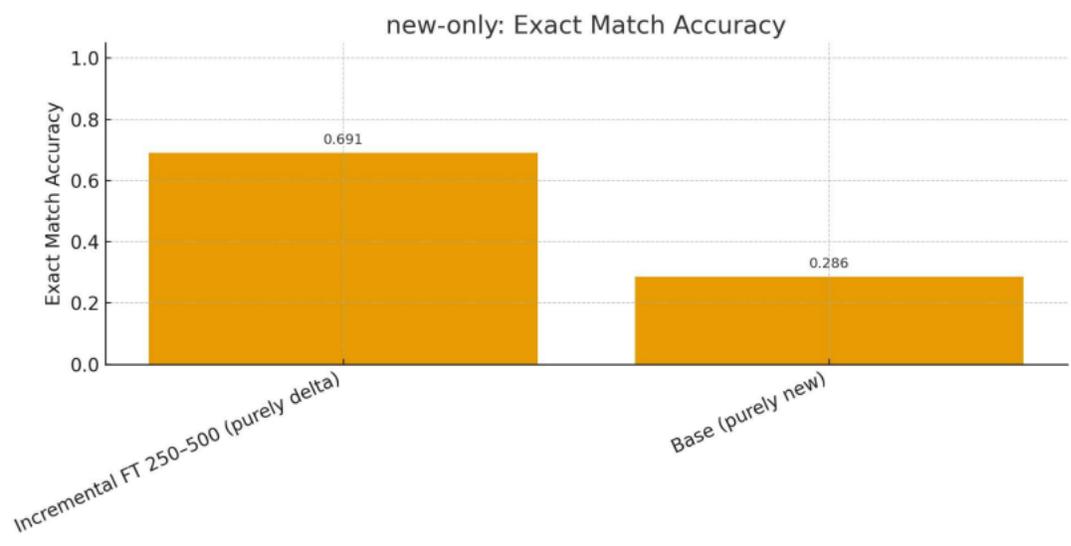

**Interpretation.** SQLAlchemy's broader windows accentuate the **prompt-budget and noise** penalties of raw diffs; **English summaries** are the right ICL form, but **Inc-FT** remains the most stable mixed-set performer.

---

# Pandas

**Setup.** Base at **100-ago**; deltas at **100→85**, **100→70**, **100→60**. We swept **NEW:OLD** mixes and **schedules (lr/epochs)** for Inc-FT and evaluated **ICL-English**.

### Range 100→85

| Variant | Old-only EM/MR | New-only EM/MR |
|---|---|---|
| **Base (100-ago)** | **0.420 / 0.420** | **0.302 / 0.302** |
| Pure NEW (350) — low lr, low ep | 0.395 | 0.240 |
| Pure NEW (350) — low lr, =ep | 0.345 | 0.292 |
| Pure NEW (350) — =lr, =ep | 0.310 | 0.240 |
| 1N:1O (350/350) | 0.370 | 0.230 |
| 1.5N:1O (525/350) | 0.302 | 0.250 |
| 1.6O:1N (438/700) | 0.330 | 0.281 |

| | | |
|---|---|---|
| 1O:1.25N (350/438) | 0.344 | 0.250 |
| 2O:1N (350/700) — low lr/ep | 0.354 | 0.230 |
| 2O:1N (350/700) — =lr/=ep | 0.210 | 0.230 |
| 3O:1N (350/1050) — lr=2.5e-5, ep=1 | 0.354 | 0.281 |
| 3O:1N (350/1050) — lr=4e-5, ep=2 | 0.330 | 0.292 |
| 4O:1N (350/1400) | 0.354 | 0.260 |
| **ICL (English)** | 0.310 | **0.830** |

**Observation.** With modest drift (100→85), the **Base** remains strongest on **OLD**, while **ICL-English** dominates **NEW** when training is skipped. Inc-FT improves NEW to ~0.292 at best, but not near ICL's NEW peak in this range.

## Range 100→70

| Variant | lr | ep | Old-only EM/MR | New-only EM/MR |
|---|---|---|---|---|
| Pure NEW (550) | 3.5e-5 | 1.5 | 0.344 | **0.330** |
| 2O:1N (1100/550) | 3.5e-5 | 1.0 | 0.333 | 0.320 |

| | | | | |
|---|---|---|---|---|
| 3O:1N (1650/550) | 3.5e-5 | 1.0 | 0.313 | 0.311 |
| 4O:1N (2200/550) | 3.5e-5 | 1.0 | **0.344** | 0.291 |
| **ICL (English)** | — | — | 0.260 | **0.864** |

**Observation.** As distance grows, **ICL-English** remains the **NEW** champion due to compact, high-signal summaries; **Inc-FT (pure NEW)** lifts NEW to ~0.33 while keeping OLD near ~0.34.

## Range 100→60

| Variant | lr | ep | Old-only EM/MR | New-only EM/MR |
|---|---|---|---|---|
| Pure NEW (750) | 4e-5 | 1.4 | 0.3125 | 0.3028 |
| 1O:1N (750/750) | 3e-5 | 1.0 | 0.3646 | 0.3303 |
| 2O:1N (1500/750) | 3e-5 | 1.0 | 0.3125 | 0.3028 |
| **2O:1N (1500/750)** | **2.5e-5** | **0.5** | **0.3646** | **0.3486** |
| 3O:1N (2250/750) | 2.5e-5 | 0.5 | 0.3542 | 0.3211 |
| 3O:1N (2250/750) | 2e-5 | 0.35 | 0.3542 | 0.3486 |
| 4O:1N (3000/750) | 2e-5 | 0.30 | 0.3646 | 0.3486 |

| | | | | |
|---|---|---|---|---|
| **ICL (English)** | — | — | 0.2100 | 0.7982 |

**Observation.** With **heavier drift**, **Inc-FT** benefits from **balanced/old-heavy mixes plus a reduced schedule** (e.g., **2O:1N, lr=2.5e-5, ep=0.5**), yielding **NEW ≈ 0.349** and **OLD ≈ 0.365**. **ICL-English** still leads NEW but does not retain OLD.

## Pandas — Comparison Charts

These charts present **Base FT vs the best Inc-FT we found vs ICL (English summaries)** for each commit range. We ran multiple mixes and hyper-parameter settings (lr/epochs) per range; the **"best Inc-FT"** is chosen by the highest EM on the relevant test split (new-only for the *new* chart, old-only for the *old* chart). See the tables above for the full sweep and exact lr/epoch settings.

### New-only (Base vs Best Inc-FT vs ICL) — EM/MR

- 100->85

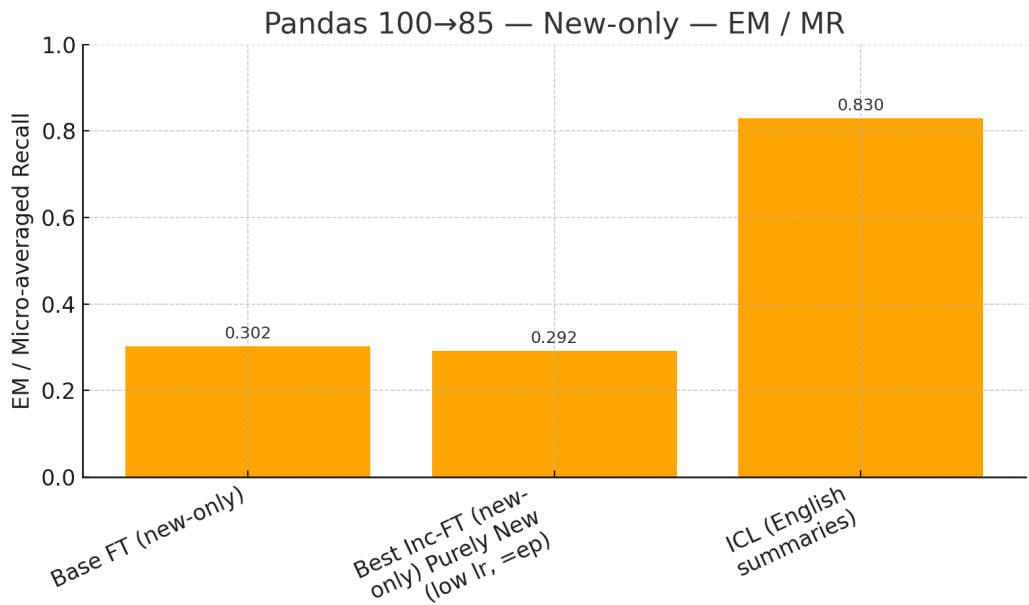

- 100->70

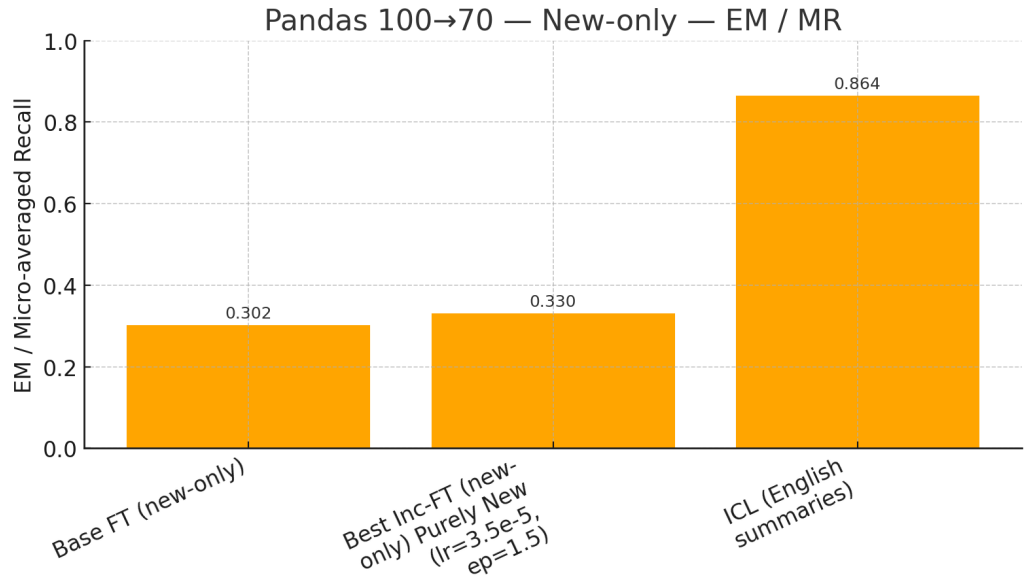

- 100->60

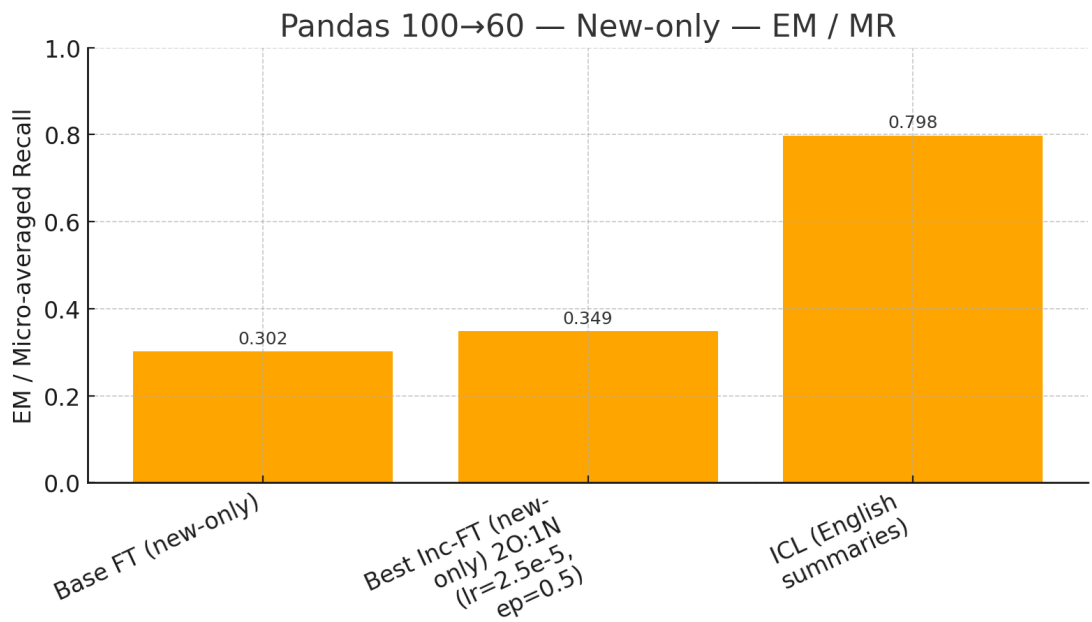

**Old-only (Base vs Best Inc-FT vs ICL) — EM/MR**
- 100->85

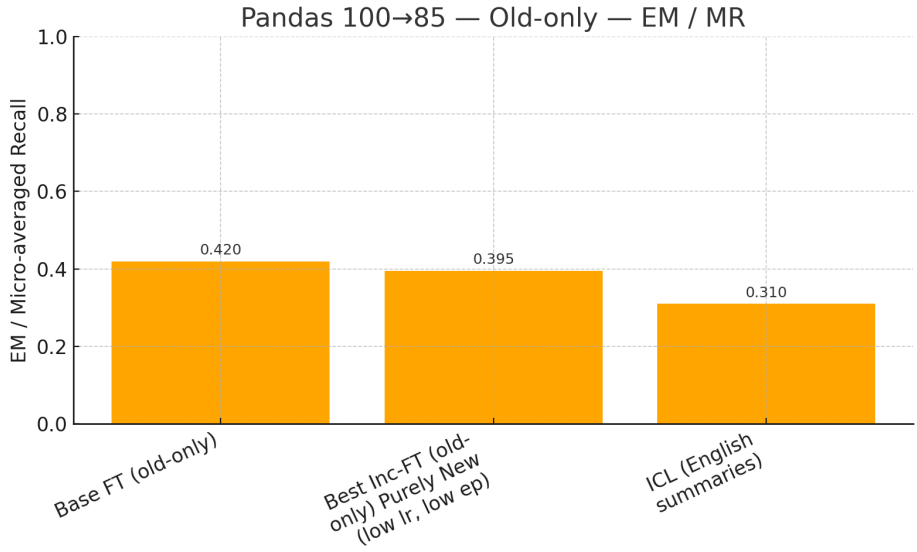

- 100->70

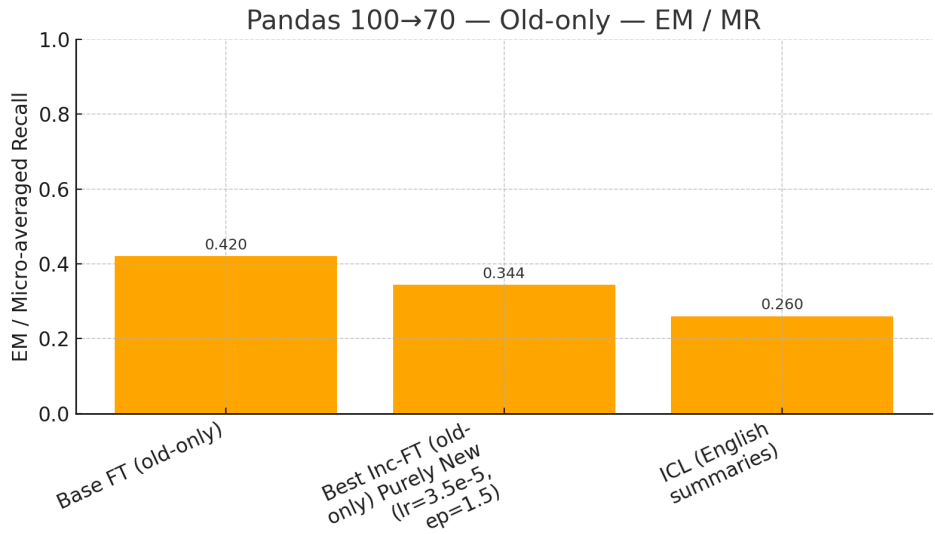

- 100>60

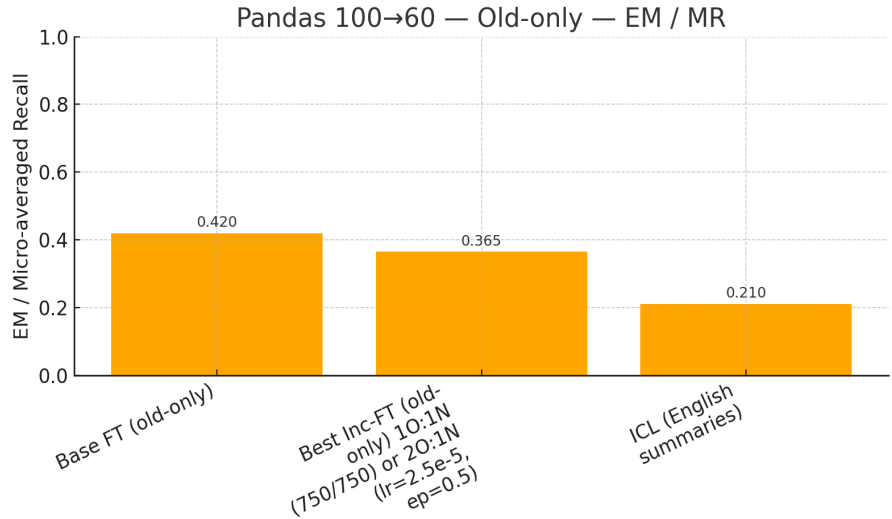

**Summary across ranges.**

- **Best Inc-FT (NEW)**: 0.292 (100→85), 0.330 (100→70), 0.349 (100→60).

- **Best Inc-FT (OLD)**: 0.395 (100→85, Pure NEW low-schedule), 0.344 (100→70), 0.365 (100→60, 2O:1N reduced schedule).

- **ICL-English** persistently dominates **NEW-only**; **Inc-FT** gives the best **mixed** trade-off with controlled forgetting.

**Note on hyper-parameter sweeps**
We experimented with multiple **mix ratios** and **training schedules (lr/epochs)** to balance *freshness* (new-only) and *retention* (old-only). Lower learning rates and shorter schedules often reduced forgetting while still capturing deltas. The charts above display the **best EM/MR** per range; tables preserve the full sweep so choices are auditable.

# Repository Case Study: Flask — "Is Inc-FT close enough to a full refresh?"

## Why this study

We want to know when **incremental finetuning (Inc-FT)** can replace a full retrain on the *latest* snapshot without losing much accuracy. Practically: if Inc-FT on a delta (x→y) is within a few points of a freshly fine-tuned model at *y*, we save time and cost.

## Concepts used here:

- **Base FT at t−x**: model trained on an older snapshot.
- **Inc-FT on (x→y) delta**: continue training the base with only the changes between t−x and t−y. (Here we keep the task fixed: predict relevant repo file paths.)
- **Cross-track comparison**:
  - **A → A\***: Base at t−x, then Inc-FT on (x→y).
  - **B**: Fresh Base FT at t−y.
  - **B\***: (For symmetry) start at t−y and also Inc-FT on the (x→y) delta; used to sanity-check schedule effects.

Evaluation: Old = "200-ago" set; New = "200→100" delta set. EM and MR track together; below we report MR (same values as EM in this task).

## Setup

- **Task**: question → relevant file path(s).
- **Data**: 820 examples to build base FTs (656 train / 164 test); delta sets for the 200→100 range.
- **Schedules**: low LR + short epochs for Inc-FT to avoid forgetting; exact values per run in tables.

**Track A (Base@200 → Inc-FT on 200->100):**

| Variant | Old MR | New MR | LR | Ep |
|---|---|---|---|---|
| A: Base@200 | 0.5427 | 0.4923 | — | — |
| A*: Pure NEW (200) | 0.5610 | 0.5231 | 3.5e-5 | 2.5 |
| A*: 1O:1N (200/200) | 0.5500 | 0.5100 | 3e-5 | 2 |
| A*: 2O:1N (200/400) | 0.5610 | 0.4308 | 2.5e-5 | 1.5 |
| A*: 3O:1N (200/600) | 0.5366 | 0.4769 | 2e-5 | 1 |

**Track A performance:**

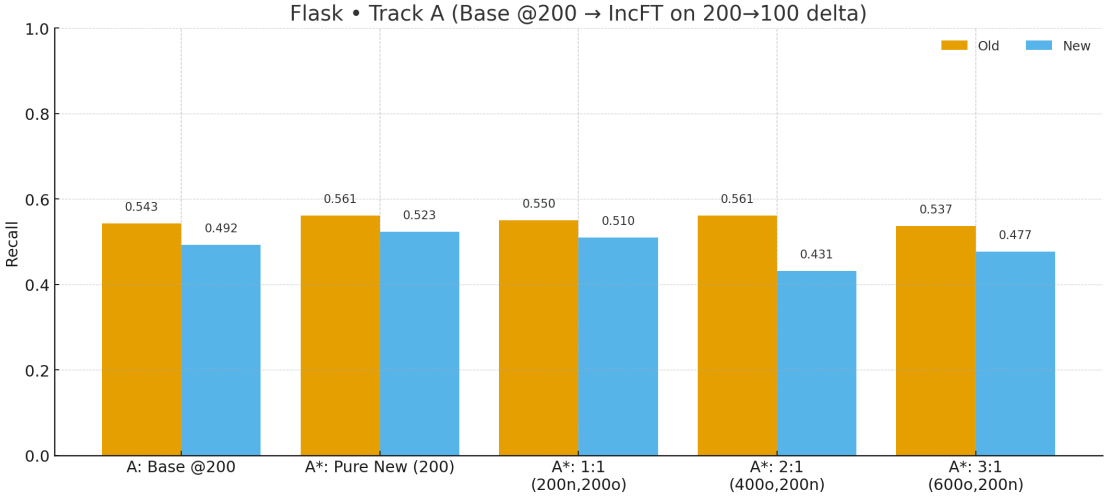

**Track A deltas vs Base:**

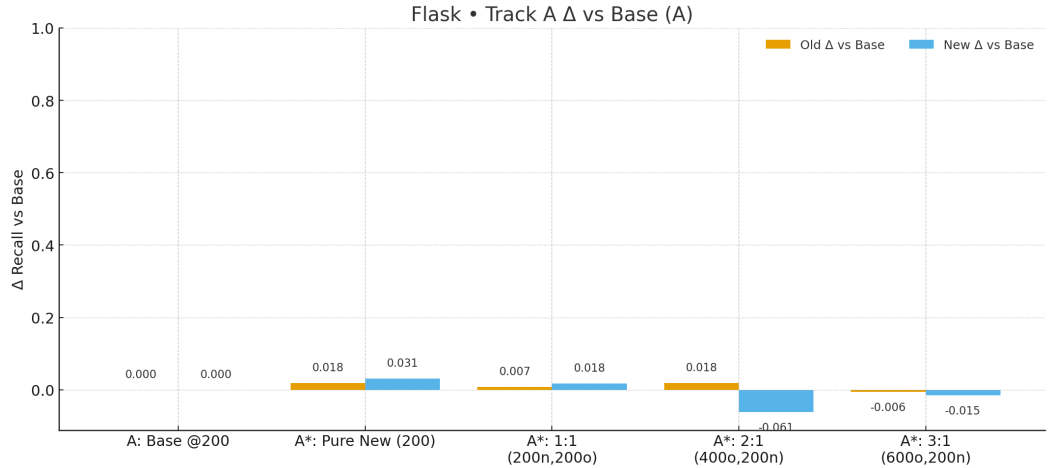

**Observations (Track A)**

- Pure-new Inc-FT slightly **beats** the base on both Old (+0.018) and New (+0.031).
- Old-heavy mixes (2O:1N, 3O:1N) **retain** old performance but can hurt New (underfitting to delta).

Track B (Base@100 → Inc-FT on the same 200->100 delta):

| Variant | Old MR | New MR | LR | Ep |
|---|---|---|---|---|
| B: Base@100 | **0.5732** | **0.5775** | — | — |
| B*: Pure NEW (200) | 0.5800 | 0.5211 | 3.5e-5 | 2.5 |
| B*: 1O:1N (200/200) | 0.5976 | 0.4789 | 3e-5 | 2 |
| B*: 2O:1N (200/400) | 0.5800 | 0.5231 | 2.5e-5 | 1.5 |
| B*: 3O:1N (200/600) | 0.5915 | 0.4930 | 2e-5 | 1 |

**Track B performance:**

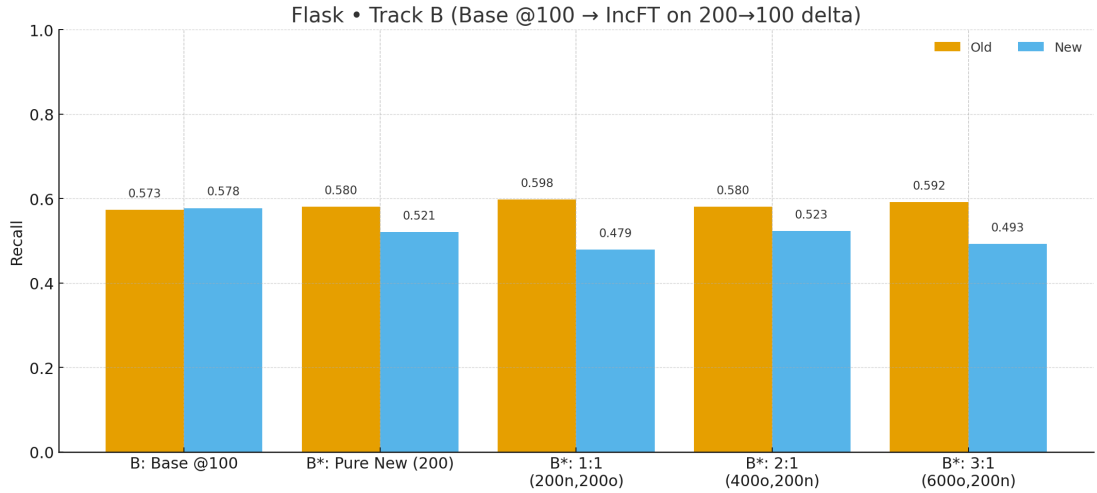

**Track B deltas vs Base:**

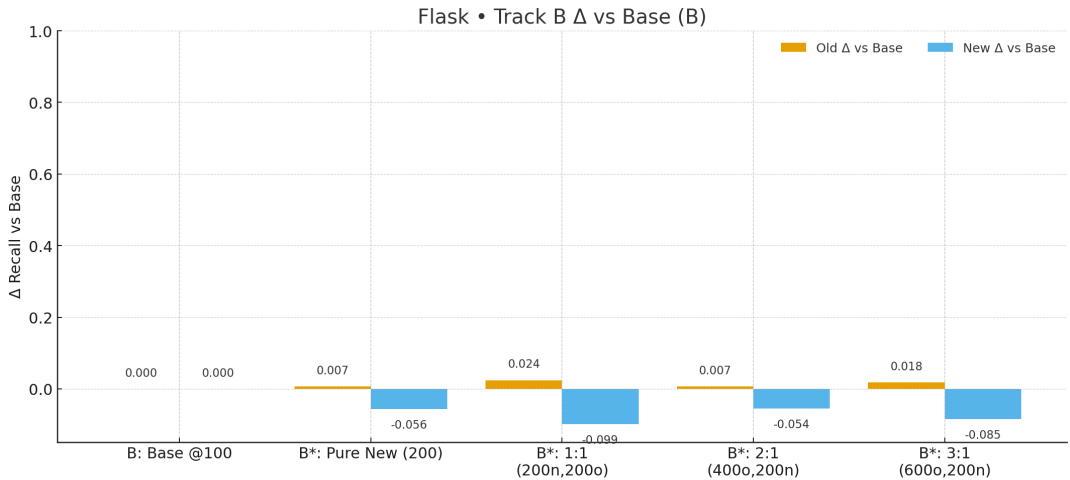

## Observations (Track B)

- The fresh base at the newer snapshot (B) is strongest on New (0.5775).
- Inc-FT from B (B*) increases Old slightly but consistently **reduces New**, especially with old-heavy mixes.

## Cross-track comparisons

- **A* vs B (Is "Inc-FT @200" ≈ "Base @100"?)**

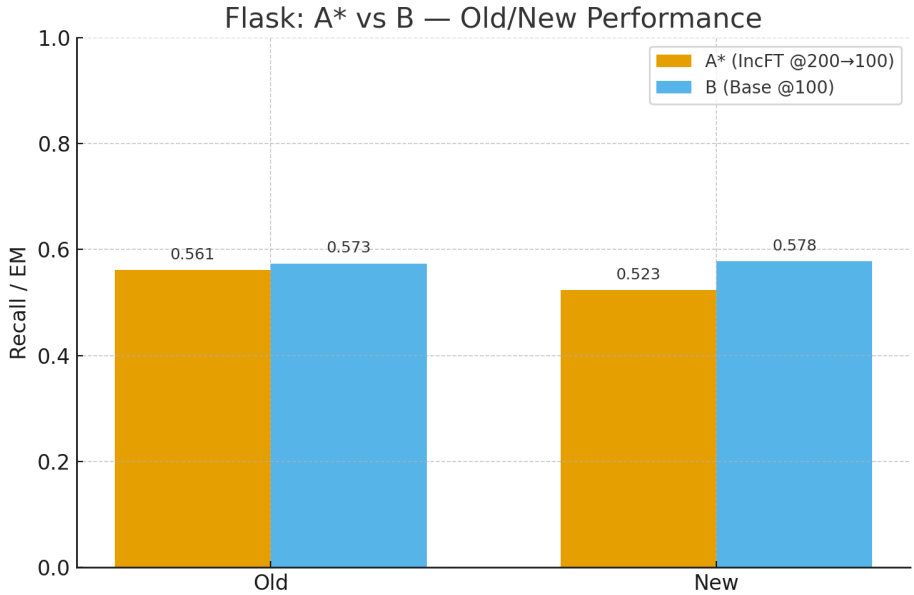

Pure-new A* reaches (Old=0.561, New=0.523) vs B (Old=0.573, New=0.578). Gap on New suggests full refresh at the later snapshot still wins when you care most about newest code.

- **A* vs B*** *(Purely-new Inc-FT on both bases)*

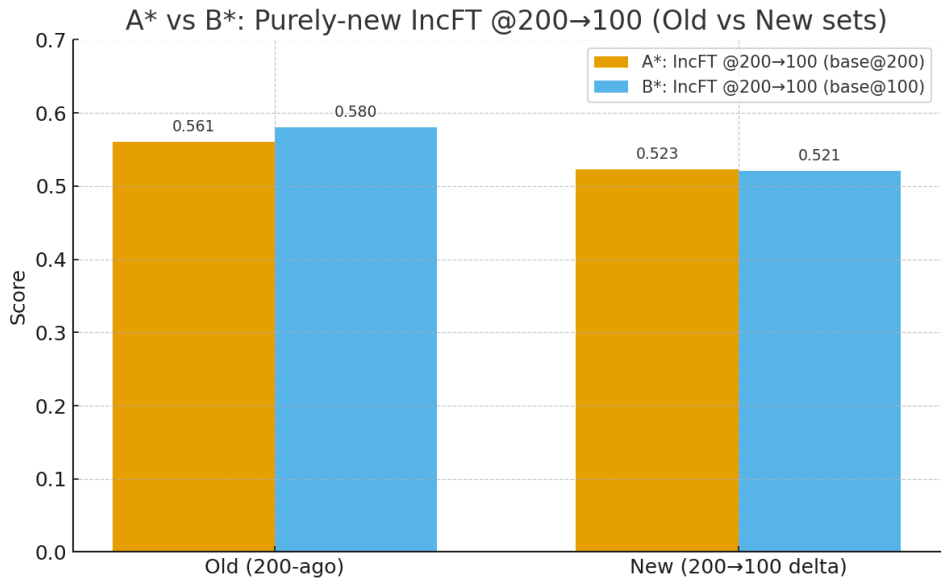

Both bases can be nudged with the same delta; the later base (B) remains ahead on New, indicating **base snapshot quality matters** as much as the delta schedule.

**Takeaways:**

---

- When New performance is critical, **full Base @latest** remains the ceiling.

- If small regressions on New are acceptable, **Inc-FT (pure-new)** on the older base delivers close-enough performance with lower cost.
- Avoid overly old-heavy mixes for deltas with substantive behavior changes—they preserve Old but can **underfit** the delta and depress New.

# Repository Case Study: Poetry — Handling Structural Drift with Alias-Aware Evaluation

## Why this study

Poetry's 1739→1639 commit window includes heavy **structural drift** (renames, deletions) alongside many **low-signal edits** (comments, type hints). This creates two challenges:

1. **Train on behavioral changes** (meaningful M edits) without ever labeling **deleted** paths as correct.
2. **Evaluate fairly** when a model still "remembers" *old* names that were **renamed or deleted**.

Our solution couples **alias-aware training data curation** with **alias-aware evaluation**.

## Experimental setup

- **Base FT @1739**: single-shot fine-tune on the older snapshot.
- **Inc-FT on 1739→1639 delta**: several OLD/NEW mixes, with NEW drawn from the 1739→1639 changes and OLD drawn from the @1739 pool (excluding items that touch M/A files at Y to avoid leakage).
- **Base FT @1639**: full refresh on the newer snapshot (a reference floor/ceiling).
- **Test splits:** 100 examples each for **Old** and **New**.
- **Metrics:** Exact Match (EM) and **Micro-averaged Recall (MR)**; we report MR/Accuracy on the charts.
- **Inference:** alias-aware parsing and scoring as described above.

## Results - MR/Accuracy (OLD vs NEW)

| Experiment | OLD MR | NEW MR |
| --- | --- | --- |

| | | |
|---|---|---|
| BASE_FT@1739 | 0.54 | 0.30 |
| **PURE_NEW_300** (LR=2e-5, Ep=0.8) | **0.53** | **0.36** |
| 2OLD:1NEW (300/600, LR=3e-5, Ep=1.2) | 0.46 | 0.32 |
| 3OLD:1NEW (300/900, LR=2.5e-5, Ep=1.3) | 0.39 | 0.35 |
| 4OLD:1NEW (300/1200, LR=2e-5, Ep=1.4) | 0.39 | 0.26 |
| BASE_FT@1639 | 0.50 | 0.35 |

*Poetry 1739→1639 — Old vs New MR/Accuracy by experiment.*

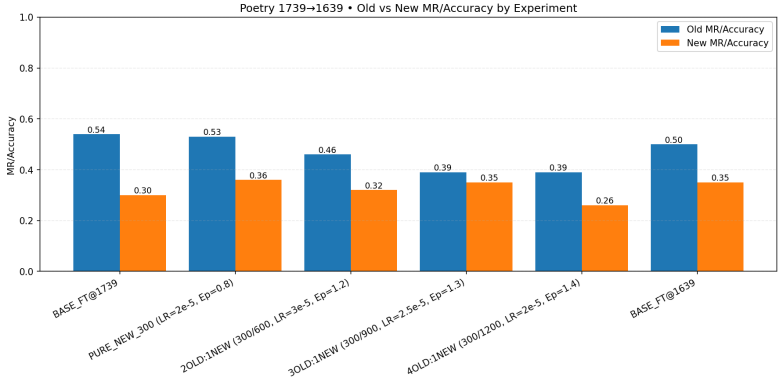

*MR/Accuracy trend across Inc-FT mixes (Old & New).*

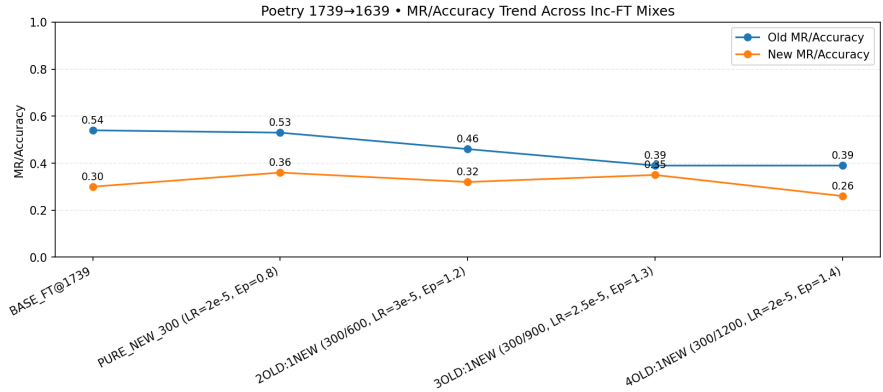

**Takeaways**

- **Pure NEW (300)** lifts **New** to 0.36 while **preserving Old** (0.53), matching/exceeding Base@1639 on New and staying close on Old.
- Heavier OLD mixing (3:1, 4:1) **suppresses New** and drags Old—consistent with older priors overwhelming the small NEW signal.
- Base@1639 is a solid refresh baseline; Inc-FT approaches it when NEW is curated and ratioed conservatively..

## Alias-aware behavior (what the model predicts)

**Alias reason breakdown (OLD split; counts out of 100)**

| Experiment | direct | alias_rename | alias_deleted | rescued_suffix | rescued_fuzzy |
|---|---|---|---|---|---|
| BASE_FT@1739 | 52 | 35 | 0 | 9 | 4 |
| PURE_NEW_300 | 53 | 36 | 0 | 7 | 4 |
| 2OLD:1NEW | 47 | 36 | 0 | 10 | 7 |
| 3OLD:1NEW | 50 | 30 | 0 | 14 | 6 |
| 4OLD:1NEW | 39 | 35 | 0 | 17 | 9 |

*Alias reason breakdown — OLD split.*

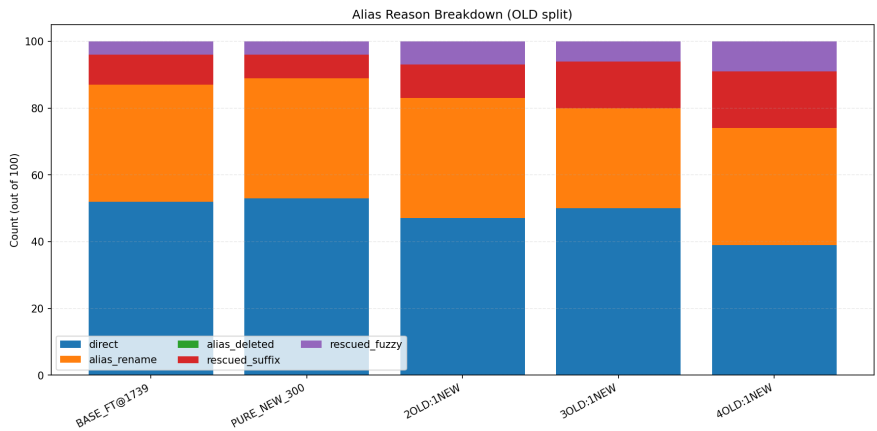

**Alias reason breakdown (NEW split; counts out of 100)**

| Experiment | direct | alias_rename | alias_deleted | rescued_suffix | rescued_fuzzy |
|---|---|---|---|---|---|
| BASE_FT@1739 | 64 | 14 | 10 | 6 | 6 |
| PURE_NEW_300 | 65 | 14 | 10 | 4 | 7 |
| 2OLD:1NEW | 65 | 11 | 10 | 6 | 8 |
| 3OLD:1NEW | 64 | 3 | 10 | 13 | 10 |
| 4OLD:1NEW | 64 | 5 | 10 | 16 | 5 |

*Alias reason breakdown — NEW split.*

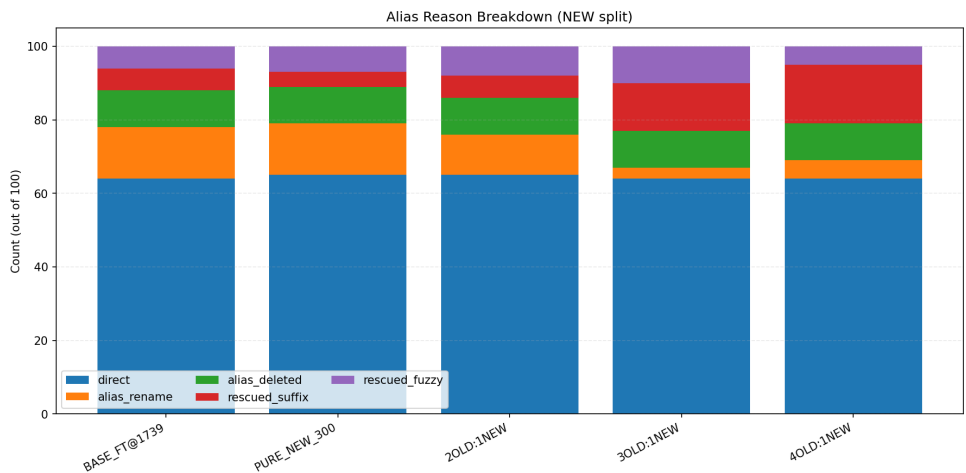

**Interpretation.** On **NEW**, we see elevated **alias_rename** and **alias_deleted** events—evidence of structural drift. The alias-aware scorer avoids penalizing correct behavior remembered under old names (renames), while **deleted** paths receive no credit but are tabulated to quantify lingering "old memory."

## In-Context vs Inc-FT vs Base

**Comparison (Old-only & New-only MR/Accuracy)**

| Setting | Old MR | New MR |
|---|---|---|
| BASE_FT@1739 | 0.54 | 0.30 |
| BASE_FT@1639 | 0.50 | 0.35 |
| INC-FT PURE_NEW_300 | 0.53 | 0.36 |
| In-Context Alias | 0.34 | 0.42 |

*Old split — In-Context vs Inc-FT vs Base.*

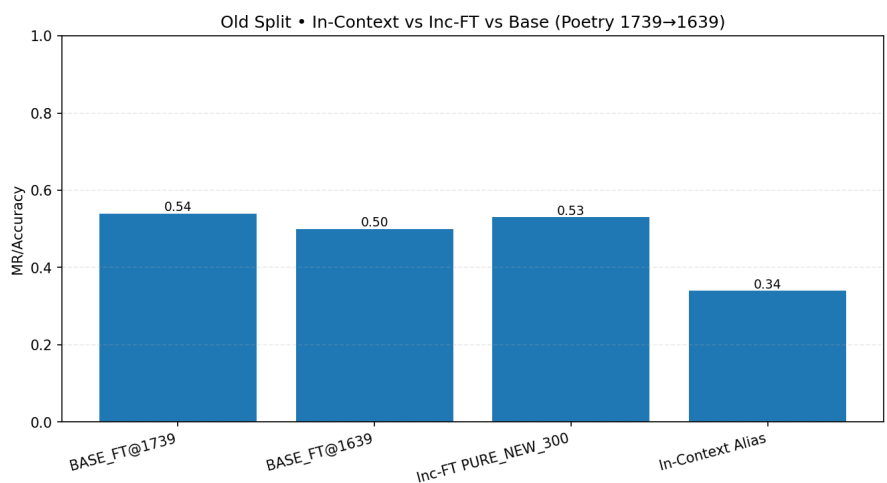

*New split — In-Context vs Inc-FT vs Base.*

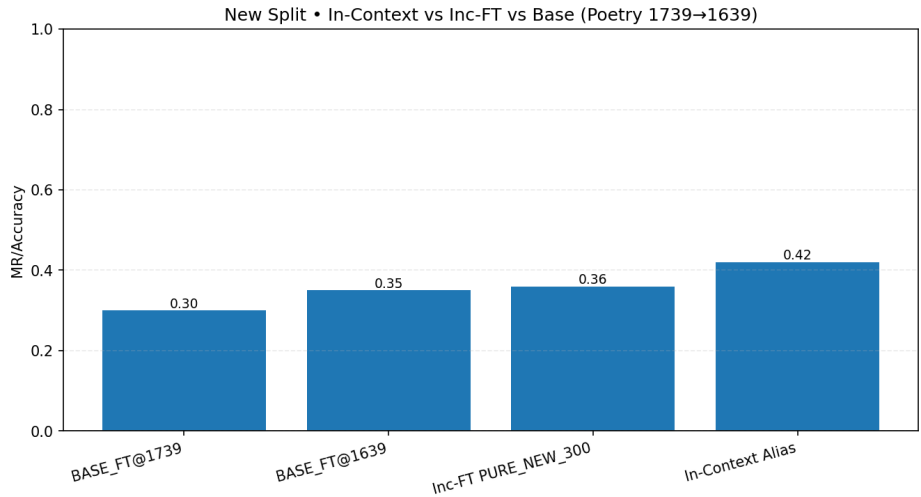

**Observations:**

- On **Old**, Pure NEW Inc-FT stays competitive with Base@1739 and Base@1639.
- On **New**, Pure NEW Inc-FT and Base@1639 sit close, both outperforming the older Base@1739—reasonable since both inject post-1739 signal (one via Inc-FT, the other via full refresh).

## Measuring forgetting explicitly (structural-change probe)

We build a **Forgetting Probe**: a test set where **every gold file** was **Renamed or Deleted** between 1739→1639. We then run a **no-remap** evaluator that counts raw emissions of **old names**.

### Forgetting probe counts (no remap; out of 100)

| old_name | new_name | deleted_old | unknown |
|----------|----------|-------------|---------|
| 78 | 0 | 0 | 22 |

### Forgetting probe rates(no remap)

| Metric | Value |
| --- | --- |
| OLD-path emission rate (% predictions that use OLD names) | 0.78 |
| EM (vs OLD names as gold) | 0.66 |
| MR (vs OLD names as gold) | 0.66 |

Forgetting Probe (Pure NEW 300): counts of old_name, new_name, deleted_old, unknown.

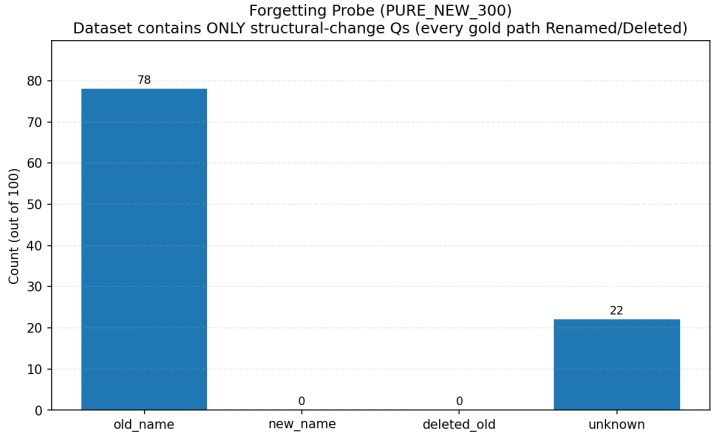

Forgetting Probe rates (Pure NEW 300). Dataset contains **only structural-change** Qs (every gold path Renamed/Deleted).

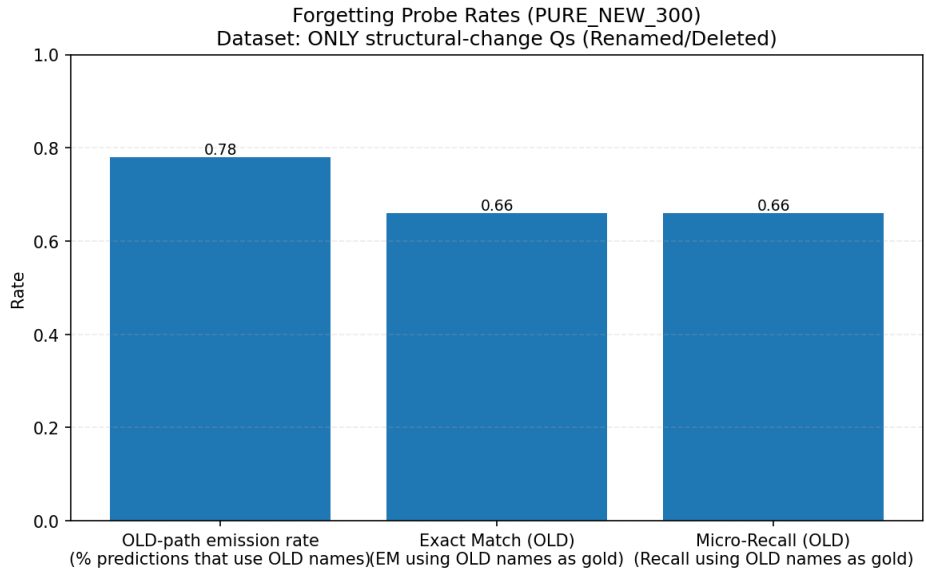

**Interpretation:**

- The model still emits many **old names** on structural-change questions, which is **expected**: Inc-FT was trained to learn **behavioral** changes from M files, not to **rewrite** structural memory.
- This probe is diagnostic: it quantifies how much "old memory" remains for **names**, independently of behavioral accuracy.

---

# Git-diff Inc-FT vs Full-file Inc-FT

We now compare the **two ways of building NEW** for Inc-FT—**Git-diff** (English delta summaries) vs **Full-file** (head-state content)—head-to-head on **Poetry** and **Flask**.

## Poetry (rename/delete-heavy)

### MR/Accuracy tables

### Side-by-side (Git-diff vs Full-file)

| Mix | Git-diff OLD | Git-diff NEW | Full-file OLD | Full-file NEW |
|---|---|---|---|---|

| | | | | |
|---|---|---|---|---|
| **Pure NEW (300)** | 0.53 | **0.36** | 0.54 | 0.30 |
| 2:1 (600 old/300 new) | 0.46 | **0.32** | 0.48 | 0.24 |
| 3:1 (900 old/300 new) | 0.39 | **0.35** | **0.49** | 0.29 |
| 4:1 (1200 old/300 new) | 0.39 | **0.26** | **0.50** | 0.25 |
| In-Context Alias (no FT) | 0.34 | **0.42** | 0.27 | 0.36 |

**Deltas (Full-file – Git-diff)**

| Mix | Δ OLD | Δ NEW |
|---|---|---|
| Pure NEW (300) | +0.01 | -0.06 |
| 2:1 (600 old/300 new) | +0.02 | -0.08 |
| 3:1 (900 old/300 new) | **+0.10** | -0.06 |
| 4:1 (1200 old/300 new) | **+0.11** | -0.01 |
| In-Context Alias (no FT) | -0.07 | -0.06 |

**Old vs New (Full-file Inc-FT only):**

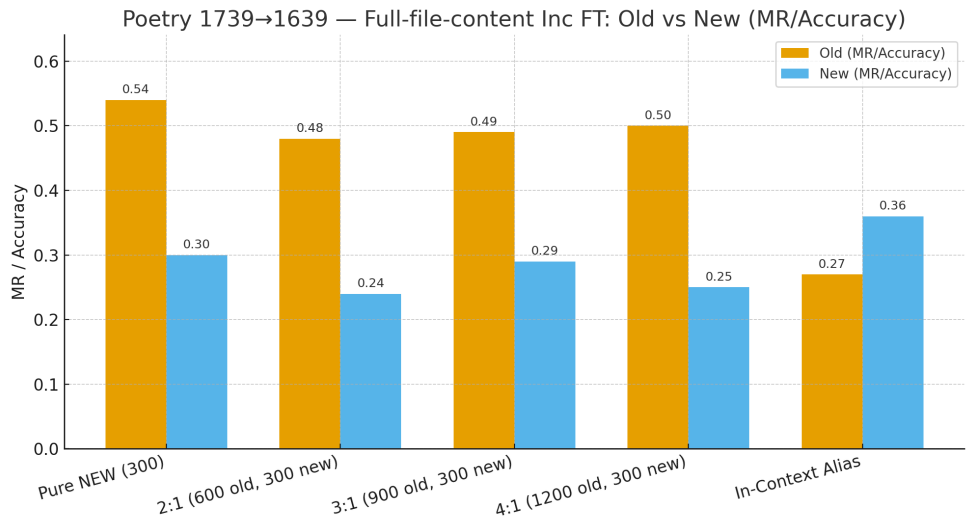

**NEW split — Git-diff vs Full-file (by mix):**

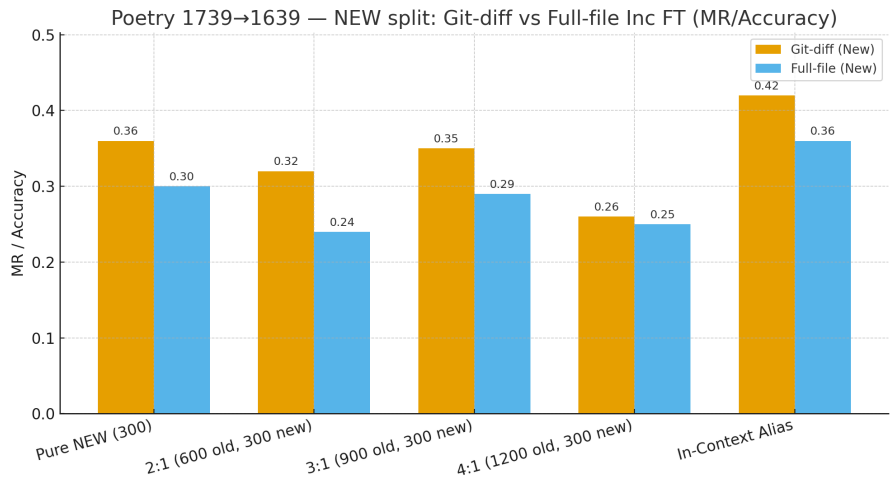

**OLD split — Git-diff vs Full-file (by mix):**

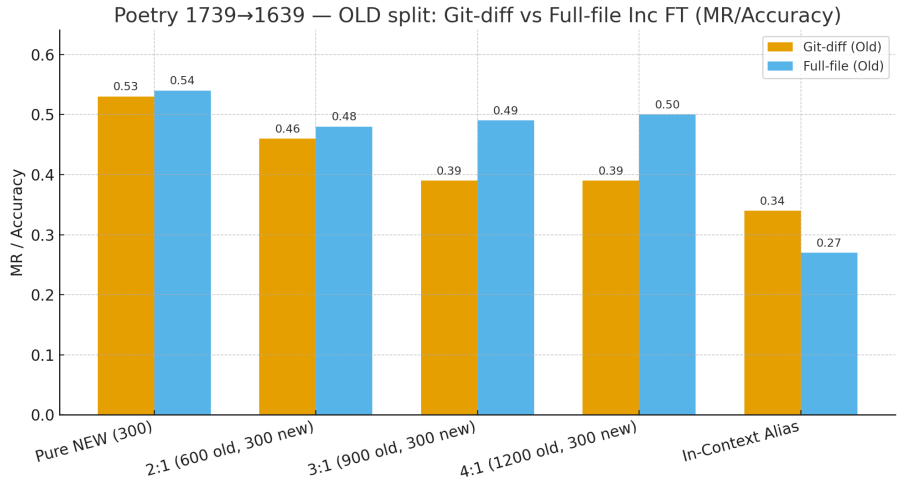

**Overview (Old & New together):**

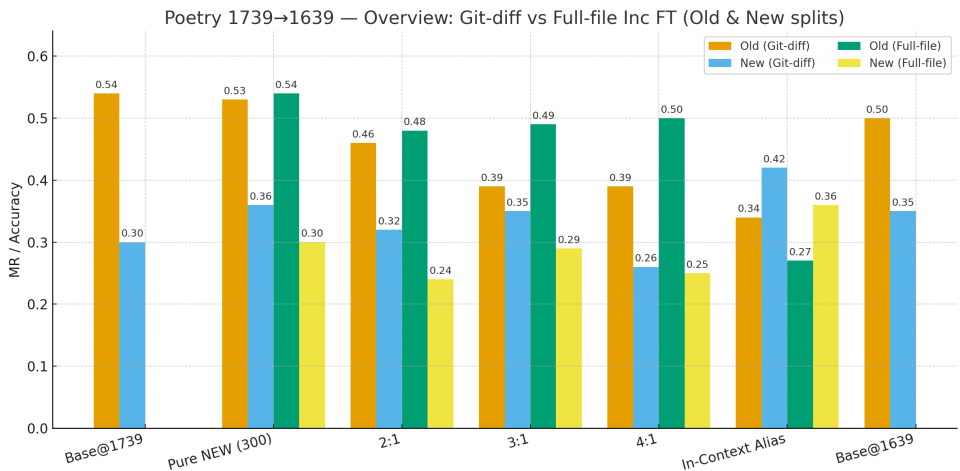

**What we see (Poetry)**

- **OLD:** Full-file improves OLD as the OLD share increases (**+0.10 to +0.11** at 3:1/4:1). Richer file context appears to stabilize legacy behavior when training is OLD-heavy.
- **NEW:** Git-diff consistently beats Full-file on this window (by 0.01–0.08). With many renames/deletes and low-signal edits, concise diff summaries keep the supervision tightly focused on the modified behavior.

**Take:** For rename/delete-heavy windows, **Git-diff** is the safer bet for **NEW**. Full-file can be used to keep **OLD** performance healthy when you run OLD-heavy mixes

# Flask (behavioral-change-heavy)

**MR/Accuracy tables**

| Mix | Git-diff OLD | Git-diff NEW | Full-file OLD | Full-file NEW |
|---|---|---|---|---|
| **Pure NEW (200)** | **0.543** | 0.523 | 0.514 | **0.643** |
| 1:1 (200 old / 200 new) | **0.543** | 0.523 | 0.530 | **0.630** |
| 300 old / 200 new | **0.514** | 0.480 | 0.470 | **0.630** |

**Deltas (Full-file – Git-diff)**

| Mix | Δ OLD | Δ NEW |
|---|---|---|
| Pure NEW (200) | -0.029 | **+0.120** |
| 1:1 (200 old / 200 new) | -0.013 | **+0.107** |
| 300 old / 200 new | -0.044 | **+0.150** |

**1:1 OLD vs NEW (Git-diff vs Full-file):**

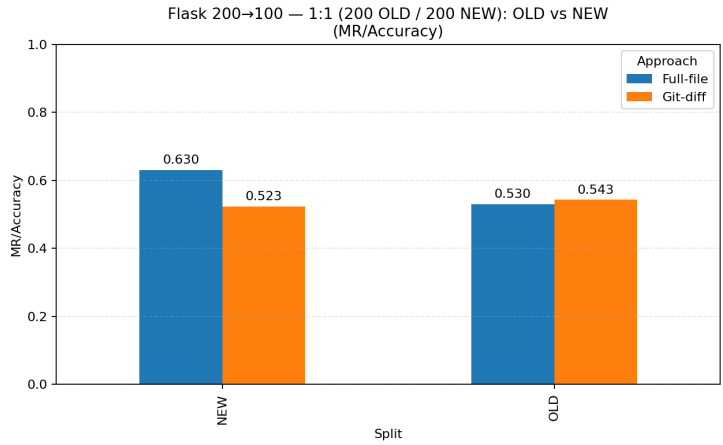

## 300 OLD / 200 NEW:

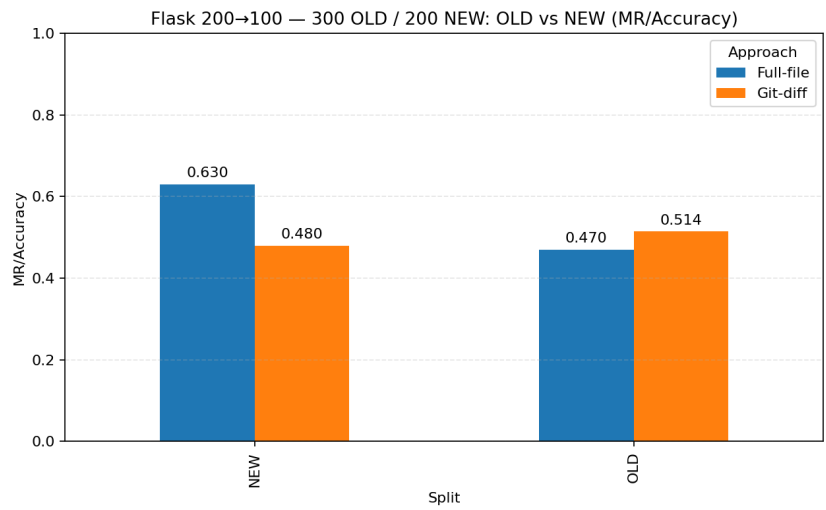

## NEW split comparison across mixes:

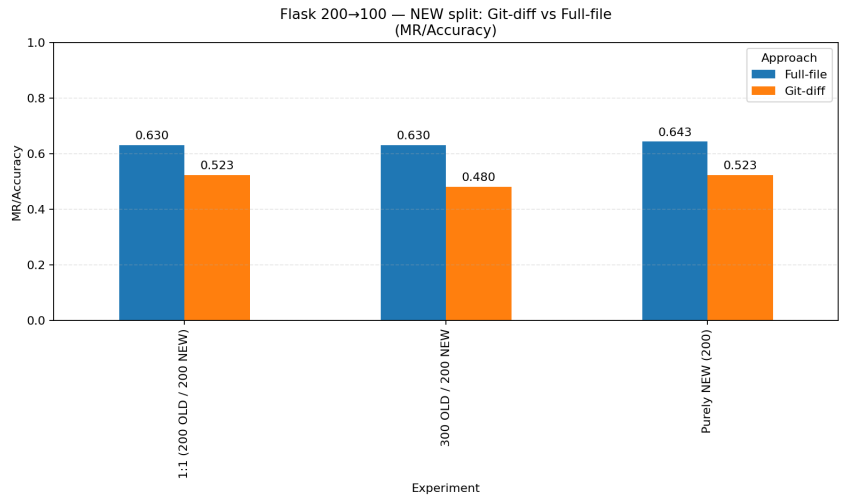

## OLD split comparison across mixes:

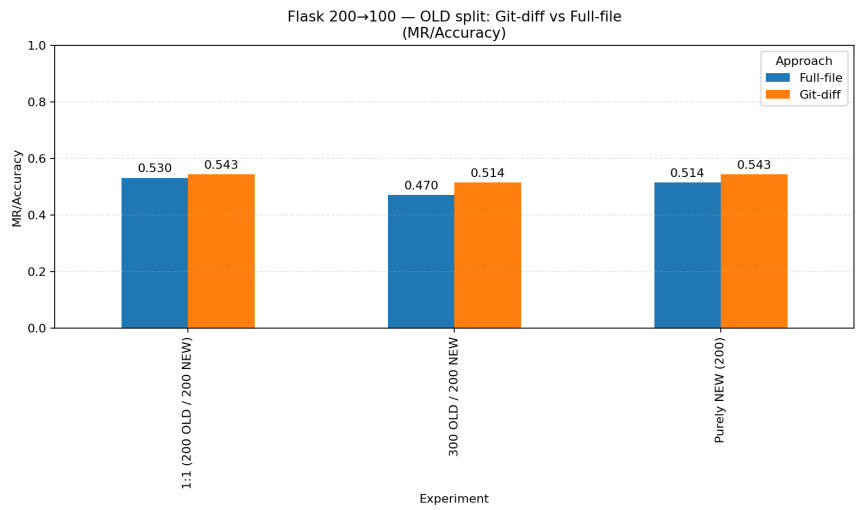

## Pure NEW (200):

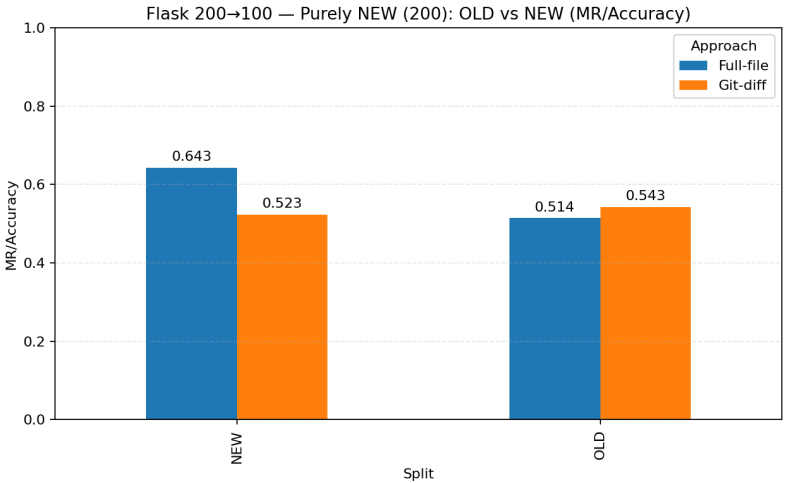

**What we see (Flask)**

- **NEW:** Full-file **wins decisively** on every mix (**+0.11 to +0.15**). When changes are behavioral and remain within files, full context teaches the model the updated logic better than a terse diff summary.
- **OLD:** Slightly behind Git-diff (−0.01 to −0.04), which is expected when you emphasize changed files over broad legacy coverage.

**Take:** For **behavioral-change-heavy** windows, **Full-file** is the clear choice to maximize **NEW** while staying close on **OLD**

# Analysis & Discussion

The results across Flask, SQLAlchemy, Pandas, and Poetry form a consistent narrative about how to keep a path-prediction model "speaking the current dialect" of a repository while not forgetting its past. Three forces recur: (i) the *character* of the commit window (behavioral changes vs. structural churn), (ii) the *vehicle* used to inject freshness (ICL with raw diffs vs. English deltas, or Inc-FT), and (iii) the *mix and schedule* choices that determine how much the model's priors are nudged or preserved. Read together, these forces explain the quantitative outcomes you saw in EM and micro-averaged recall, and they motivate the practical rules of thumb we propose at the end of this section.

# 1) What the numbers say—one story across repos

Start with Flask, where the window is relatively behavioral and compact (≈10 commits). Here, the incremental fine-tuning setting that mixes *new* with a modest surplus of *old* (96n/192o) consistently produces the best mixed and new-only performance. This pattern is telling: even when the supervision is aimed at changed files, a measured replay of legacy examples acts like a stabilizer, preventing the adapter from overfitting the surface of the diff while helping it internalize invariants that still govern the codebase. In contrast, the "balanced" 96n/96o regime underperforms dramatically; it doesn't deliver enough gradient on the new behavior to overcome drift, yet it still perturbs the base distribution enough to forget useful priors. ICL also helps here, and—importantly—*diff-only* ICL is competitive in Flask because the diffs are short, mostly behavioral, and therefore semantically dense; the prompt has headroom to carry raw patches without drowning in churn.

Shift to SQLAlchemy and Pandas, where commit distance grows and diffs are longer and noisier. The headline is that English delta summaries dominate raw diffs for ICL as distance increases. This is exactly what we would expect when prompt entropy rises: raw technical patches accumulate formatting, typing, and comment changes that add tokens without adding supervision. Compressing those patches into short English descriptions filters out low-signal edits and foregrounds the operational deltas ("guard added before iterating; changed default enum handling"), letting the base model attend to *what changed* rather than *how it changed on disk*. Even there, however, Inc-FT tends to win on *mixed* sets because it rewrites the model's internal priors instead of renting accuracy from the prompt window; the price is training time, but the payoff is persistent gains that don't consume inference context.

Finally, consider Poetry, the most structurally turbulent window (renames/deletions plus low-signal churn). Here, the alias-aware evaluation becomes the difference between a fair test and a misleading one. Without remapping, the model would be penalized for "remembering the right file under the wrong name," a failure that reflects structural drift rather than behavioral misunderstanding. With remapping, you see a clean split: alias_rename events are credited (because behavior lives on under a new path), alias_deleted events are not (because behavior truly disappeared), and rescued_* events are tracked to show how often near-misses get corrected. In this regime, Git-diff Inc-FT beats Full-file Inc-FT on new-only because concise deltas keep the training signal tightly coupled to the changed behavior, whereas full-file supervision admits more legacy detail per example and dilutes the gradient. That same richness, however, helps Full-file in old-heavy mixes by stabilizing legacy behavior—precisely the OLD lift we observed at 3:1 and 4:1.

# 2) Why these patterns emerge—mechanisms, not just metrics

Two mechanisms recur in error reviews and ablations:

**Prompt entropy vs. semantic compression (ICL).** As windows widen, raw diffs inflate context with edits that rarely change semantics. English deltas act as a lossy compressor tuned for

supervision: they erase layout and typing churn while preserving symbol names, control-flow changes, and guard conditions—exactly what teaches the model how to update its internal associations ("questions mentioning X now relate to Y' instead of Y"). This is why ICL-English overtakes ICL-diff in SQLAlchemy and Pandas, and why it posts very large NEW-only MR in Pandas (≈0.80–0.86) even when Inc-FT has not been run.

**Signal dilution vs. retention (Inc-FT).** Inc-FT must trade off moving toward Y while maintaining competence on X. Too little NEW or too cool a schedule underfits the delta; too much heat or too NEW-heavy a mix erases useful priors. The winning points balance these: in Flask, the old-leaning 96n/192o mix gives the adapter enough NEW to learn changed behavior while the OLD replay acts as weight decay on forgotten competencies. In Pandas 100→60, the best joint frontier emerges with slightly old-heavy mixes and *short* schedules (e.g., 2O:1N with reduced LR/epochs), which curb forgetting while still lifting NEW to the mid-0.3s MR. These combinations aren't magic numbers; they are evidence that the *shape* of the supervision (Git-diff vs. Full-file) and the *temperature* of training (LR×epochs) must be matched to the window.

**Alias awareness as instrumentation, not decoration.** In Poetry, the forgetting probe (no remap) shows high old-name emission rates. That does not mean the model failed to learn new behavior; it means the model retained surface strings for objects whose behavior migrated. Alias-aware scoring prevents conflating string memory with functional understanding, and the reason logs (direct vs alias_rename vs alias_deleted vs rescued_*) make improvements legible. When your metric moves because alias_rename rescues increased, you learned *structure*; when direct hits rise on new-only while alias_deleted stays flat, you learned *behavior*. This instrumentation is why we can diagnose windows where training helped, but old names still dominate the model's *lexicon*—a subtle but actionable finding.

## 3) Git-diff vs. Full-file Inc-FT—when each supervision mode wins

The head-to-heads in Poetry and Flask make the selection rule crisp:

- **Rename/delete-heavy or low-signal windows → Git-diff Inc-FT.** In Poetry, Git-diff consistently beats Full-file on NEW by 1–8 MR points at equal mixes because its supervision is precisely where behavior changed. Full-file injects lots of unchanged context and, in NEW-limited training budgets, that extra context steals capacity from the delta.
- **Behavioral windows with localized logic changes → Full-file Inc-FT.** Flask is the mirror image: Full-file wins NEW decisively (+0.11 to +0.15 MR) across mixes, because understanding a new guard or widened invariant often requires reading the *entire* class/module, not just the changed lines. The small OLD dip (≈1–4 points) is the predictable tax for emphasizing changed files over broad replay; when OLD is paramount, you can bleed the schedule or tilt the mix slightly towards OLD to cushion the dip.

This is not an aesthetic preference; it is a bias-variance trade-off. Git-diff reduces variance in supervision by focusing the gradient on exactly the edited symbols; Full-file reduces bias by

providing the broader context necessary to learn cross-method invariants. The window's character decides which error dominates.

|  | Git-diff | Full-file |
| --- | --- | --- |
| Rename/Delete-heavy | Best | OK |
| Behavioral change within files | OK | Best |

## 4) Full refresh vs. Inc-FT—how close is "close enough"?

The Flask A/B study answers a practical question: can Inc-FT on a delta substitute for a full refresh at the newer snapshot? The answer is "sometimes, with caveats." Pure-NEW Inc-FT on the older base narrows the gap on both OLD and NEW, but a fresh base at the newer snapshot still sets the ceiling on NEW. This is intuitive: base pretraining on the current distribution gives you better priors *everywhere*, whereas an adapter trained only on changed files cannot fully rewrite stale associations in untouched areas. If the target is "good enough now with low cost," Inc-FT does the job; if the target is "best possible NEW," full refresh remains the benchmark.

## 5) Mix and schedule—what actually moves EM/MR

Across repos, two levers repeatedly shift the frontier:

- **Mix (NEW:OLD).** Too balanced can be worst of both worlds (Flask 96n/96o); slightly OLD-leaning helps on mixed sets when windows are not extremely behavioral; NEW-heavy helps when the base is already recent and you're targeting NEW. The right point is window-dependent, but harmful extremes are consistent: heavy OLD in behavioral windows depresses NEW; pure NEW in rename/delete-heavy windows can lift NEW but risks brittle retention unless schedules are cooled.
- **Schedule (LR×epochs).** Shorter, cooler schedules reduce forgetting for a given mix. In Pandas 100→60, reducing LR/epochs on 2O:1N turns a middling setting into the best joint OLD/NEW performer. You can think of schedule as a *global regularizer* on the adapter; if you tilt NEW upward in the mix, you likely need to cool the schedule to keep OLD intact.

| Max NEW |
| :---: |
| Pure NEW or 1:1 |
| Short epochs |
| Mid/low LR |

| Balance |
| :---: |
| <= 2:1 OLD:NEW |
| Short epochs |
| Lower LR |

## 6) The role of ICL in the toolkit

ICL is not just a stopgap; it is a fast and robust knob for injecting freshness, especially as distance grows and for scenarios where training cadence is constrained. The data show a simple rule: *use English deltas* unless diffs are tiny and crisp (Flask), in which case raw diffs can be fine. ICL lifts NEW now, requires no weight updates, and—when paired with alias-aware evaluation—provides an immediate path to correctness without rewarding deleted paths. Its main trade-offs are prompt cost and the fact that it does not rewrite priors, which is why it trails Inc-FT on mixed sets once training is on the table.

# Final Recommendations

## 1) Choose the update mode by budget and drift

- **Small to moderate drift and training constrained:** *ICL with English delta summaries*, scored with the alias-aware evaluator. Expect large NEW-only gains; mixed improvements depend on prompt budget.
  *Why:* Semantic compression beats prompt entropy; no forgetting risk because weights are untouched.
- **Sustained updates with modest training budget:** *Inc-FT on a delta dataset*, pick supervision by window character:
  – *Rename/delete-heavy or low-signal churn:* **Git-diff Inc-FT**. Slightly OLD-leaning mix (≤1N:2O) + short, cool schedule to preserve OLD while learning the delta.
  – *Behavioral, within-file changes:* **Full-file Inc-FT**. NEW-balanced or pure NEW mix + short schedule to maximize NEW without large OLD regression.
- **Large accumulated drift or release-scale refactors:** *Full refresh @ HEAD* if latency/cost allow; this resets priors and sets the NEW ceiling. If timelines are tight, run a short Inc-FT bridge (per the rules above) while the full refresh trains.

| | Small Drift | Moderate-Large Drift |
|---|---|---|
| Limited Budget | ICL (English) + Alias-aware Eval | ICL or Full Refresh depending on SLA |
| Training Available | Inc-FT (Git-diff or Full-file based on delta) | Full Refresh @ HEAD |

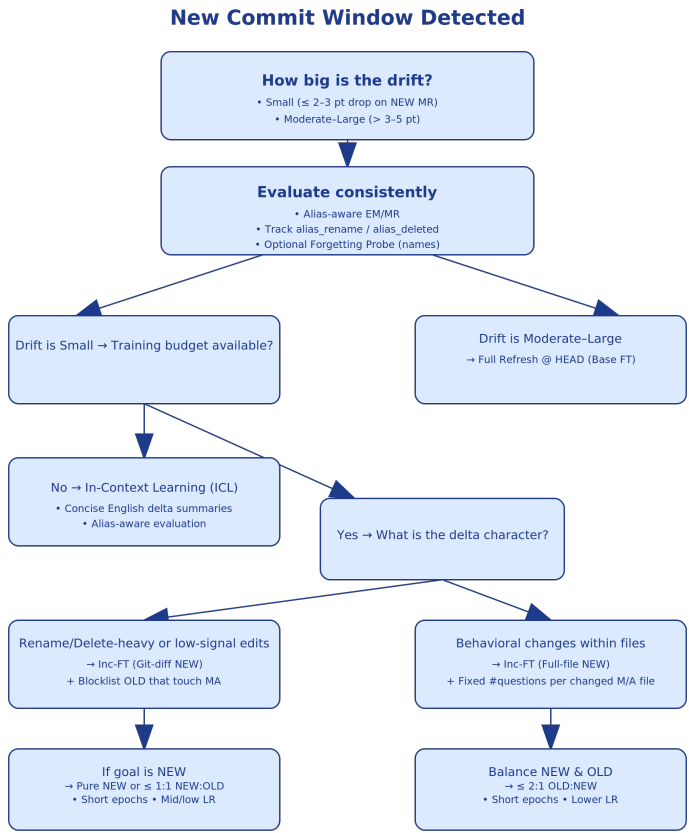

## 2) Always evaluate alias-aware—and instrument the reasons

- Score EM/MR after mapping old→new with the alias map; never credit deleted paths.
- Log `alias_rename`, `alias_deleted`, `rescued_suffix`, `rescued_fuzzy` to separate structural rescues from behavioral learning and to diagnose residual "old dialect" emissions.
- Periodically run the *Forgetting Probe* (no remap) on a structural-change slice to quantify raw old-name usage; treat a high rate as a *diagnostic* for structural memory, not as a failure of behavioral learning.

## 3) Set mix & schedule with intent

- If the base is recent and the window is behavioral: start at *pure NEW* or *1:1*, short epochs, mid/low LR.
- If the base is older or the window is churny: start at *≤1N:2O* with a *cool* schedule; increase NEW only if NEW-only MR stagnates.
- Avoid "default balance" (e.g., 1:1) as a reflex; your best point depends on the window's *signal-to-churn* ratio, not symmetry.

## Threats to Validity

- **Window heterogeneity.** Extremely large monorepos or windows with cascading cross-file effects may need hierarchical NEW construction and heavier OLD replay than we tested.
- **Summary faithfulness.** ICL and Git-diff Inc-FT rely on accurate English deltas. We mitigate speculation through deterministic prompts, but errors will attenuate gains.
- **Statistical confidence.** On small splits (e.g., Flask NEW-only=24) prefer paired bootstraps and avoid over-interpreting single-point deltas within CI.

# Limitations and Future Work

Our study demonstrates that an alias-aware formulation plus a portfolio of update strategies (Full Refresh, ICL, and Incremental FT) can keep a repository-bound retriever aligned with a moving codebase. At the same time, several limitations qualify the scope of our claims and suggest concrete extensions.

**Repository specificity and portability.** Each trained model is tuned to a single repository snapshot or delta window. This matches the operational reality we target (teams update one codebase at a time), but it limits portability: a fleet of repos implies a fleet of adapters and window bundles. Multi-repo training with explicit repository identifiers and per-repo path vocabularies is a plausible extension, but introduces a much larger label universe and harder disambiguation. A realistic compromise is *lightweight per-repo adapters* on top of a shared backbone plus an automatic "which repo and window?" router.

**Dependence on Git metadata and aliasing quality.** Our evaluation—and some safety rails during data curation—lean on a version-aware alias map (old→new or old→`__DELETED__`). This works well for standard `R/M/A/D` events, but is brittle when files split/merge, when code moves across packages with heavy rewriting, or when non-file resources (schemas, configs) encode behavior without clean Git signals. Future work should (i) fuse content hashes and fuzzy structure signatures more aggressively, (ii) attach confidence to alias decisions, and (iii) surface unresolved cases rather than silently mapping.

**Quality of delta supervision.** For ICL and Git-diff Inc-FT, English summaries of diffs are the critical signal. Summarizers can under- or over-specify behavior, and long, low-signal diffs dilute

prompts or training batches. Our safeguards (tight reviewer-style prompting; "no functional change" notices) help but do not eliminate noise. A natural extension is *learned summarization tuned against downstream retrieval loss* and automatic detection of noise-dominated diffs that should be omitted or down-weighted.

**Forgetting versus underfitting trade-offs.** Inc-FT hinges on NEW:OLD mix and schedule. Although our playbook generalizes across repositories, the sweet spot is still window-specific and sensitive to LR/epoch choices. Under-balanced mixes can either forget useful history (too NEW-heavy) or underfit the delta (too OLD-heavy). We currently choose ratios by sweep; future work should *learn the mixer*—for example, a Bayesian controller that uses short pilot runs to pick ratios and schedules under a compute budget.

**Metric coverage.** EM and micro-averaged recall (MR) are principled for set-valued retrieval, and our alias-aware scorer corrects structural drift. Still, they miss developer-centric utility (e.g., "one essential file plus a harmless extra" vs "missed the only essential file"). A richer evaluation should add *per-instance utility curves*, calibration of set size, and time-to-first-useful-file in user studies.

**ICL operational constraints.** ICL is the fastest path to freshness, especially with English deltas, but it carries prompt-size and latency costs and depends on accurate selection of which deltas to show. Large windows push against model context limits. A practical next step is a *retrieval-gated ICL* that selects only the smallest set of deltas predictive for a given question, or a two-stage scheme where an Inc-FT adapter provides a shortlist that ICL refines.

**Language and artifact coverage.** Experiments focus on Python-centric repositories. Multi-language repos, generated code, or non-code artifacts (migrations, CI config) introduce different drift patterns. Extending the delta constructors to Tree-sitter-based multi-language parsing and teaching the alias/index machinery to reason over mixed assets is straightforward engineering but untested in our study.

**Security & privacy.** When repositories are private, diff summarization and data synthesis must run on trusted infrastructure; sending raw diffs to third-party services may be unacceptable. Our method is compatible with self-hosted models, but operational guidance for secure deployments is outside scope.

**Release and reproducibility.** We make heavy use of *window bundles* (manifests, alias maps, delta stores, mix recipes). Releasing anonymized bundles and harness code would aid replication; curating such a corpus across popular repos is part of our planned work.

*Future Work.* Beyond the points above, we see three promising directions: (i) **Adaptive strategy selection**—a learned policy that observes measured drift (drop on NEW MR, alias_deleted frequency, diff density) and chooses ICL vs. Inc-FT vs. Full Refresh automatically; (ii) **Continual learning regularizers**—combining Inc-FT with parameter-importance methods (e.g., EWC/L2-SP) and selective replay to further suppress forgetting without increasing OLD

volume; (iii) **Knowledge distillation across strategies**—periodically distilling an ICL-enhanced teacher into a compact Inc-FT student to amortize prompt cost while retaining freshness.

# Conclusion

We formalized repository question-to-paths retrieval as **set-valued prediction over a moving label space** and showed that naive scoring breaks under structural drift. Our alias-aware index and scorer separate behavior from structure, crediting renames while never rewarding deleted paths. Building on this, we compared three practical update levers: **Full Refresh** (ceiling on NEW, highest cost), **ICL** (English deltas > raw diffs as distance grows; fastest to ship), and **Incremental Fine-Tuning** (persistent lift with low runtime cost when mixes and schedules are chosen well). Across four representative repositories, the pattern is consistent: **Inc-FT with OLD-aware mixing** offers the best balance on mixed sets and retains historical competence; **ICL with English deltas** is the right tool for immediate freshness; **Full Refresh @ HEAD** remains the accuracy ceiling when drift is large or structural. We distilled these findings into a **decision playbook**: when to choose Git-diff vs. Full-file supervision, how to ratio NEW and OLD, and how to set gentle schedules that trade lift for retention responsibly.

Methodologically, our contributions are (i) a clean task formulation and output contract that makes evaluation auditable; (ii) a **version-aware aliasing and scoring** stack that generalizes across refactor-heavy windows; (iii) **delta-aware data construction** in two modes (Git-diff and Full-file) plus a principled mixing scheme; and (iv) **operational guidance** that translates experimental curves into day-to-day choices for teams. The empirical evidence suggests that small, well-targeted updates—either as adapters or as prompts—are sufficient to keep a parametric index in sync with fast-moving code, provided we respect the boundary between structural drift and behavioral change. We hope these ideas—particularly alias-aware evaluation and the mix-and-schedule playbook—serve as practical building blocks for robust, low-friction maintenance of repository-aware LLMs.